\documentclass[reprint, prresearch, amsmath,amssymb,
 aps,superscriptaddress
]{revtex4-2}

\usepackage{dcolumn}
\usepackage{bm}
\usepackage{graphicx}          

\usepackage{hyperref}
\usepackage{float}

\usepackage{braket} 
\usepackage[english]{babel}    
\usepackage[normalem]{ulem}

\usepackage[dvipsnames]{xcolor}
\usepackage{zx-calculus}

\newcommand{\kket}[1]{|#1 \rangle \rangle}
\newcommand{\bbra}[1]{ \langle \langle #1 |}
\newcommand{\bbrakket}[2]{\langle \langle #1 | #2 \rangle \rangle}

\usepackage{todonotes}

\definecolor{colorZxZ}{RGB}{255,255,255}
\definecolor{colorZxX}{RGB}{204,204,204}

\begin{document}

\title{Bridging wire and gate cutting with ZX-calculus}

\author{M. Schumann}
\email[Author mail: ]{ma.schumann@fz-juelich.de}
\affiliation{Forschungszentrum J\"ulich GmbH, Peter Gr\"unberg Institute, Quantum Computing Analytics (PGI-12), 52425 J\"ulich, Germany }
\affiliation{Theoretical Physics, Saarland University, 66123 Saarbr\"ucken, Germany }
\author{ T. Stollenwerk}
\affiliation{Forschungszentrum J\"ulich GmbH, Peter Gr\"unberg Institute, Quantum Computing Analytics (PGI-12), 52425 J\"ulich, Germany }
\author{ A. Ciani}
\affiliation{Forschungszentrum J\"ulich GmbH, Peter Gr\"unberg Institute, Quantum Computing Analytics (PGI-12), 52425 J\"ulich, Germany }

\begin{abstract}
Quantum circuit cutting refers to a series of techniques that allow one to partition a quantum computation on a large quantum computer into several quantum computations on smaller devices. This usually comes at the price of a sampling overhead, that is quantified by the $1$-norm of the associated decomposition. The applicability of these techniques relies on the possibility of finding decompositions of the ideal, global unitaries into quantum operations that can be simulated onto each sub-register, which should ideally minimize the $1$-norm. In this work, we show how these decompositions can be obtained diagrammatically using ZX-calculus expanding on the work of Ref.~\cite{ufrecht2023}. The central idea of our work is that since in ZX-calculus only connectivity matters, it should be possible to cut wires in ZX-diagrams by inserting known decompositions of the identity in standard quantum circuits. We show how, using this basic idea, many of the gate decompositions known in the literature can be re-interpreted as an instance of wire cuts in ZX-diagrams. Furthermore, we obtain improved decompositions for multi-qubit controlled-Z (MCZ) gates with $1$-norm equal to $3$ for any number of qubits and any partition, which is known to be optimal. This decomposition uses a single ancilla qubit. Our work gives new ways of thinking about circuit cutting that can be particularly valuable for finding decompositions of large unitary gates. Besides, it sheds light on the question of why exploiting classical communication decreases the 1-norm of a wire cut but does not do so for certain gate decompositions. In particular, using wire cuts with classical communication, we obtain gate decompositions that do not require classical communication. 
\end{abstract}

\maketitle

\section{Introduction}
\label{sec:intro}
Quantum computers have witnessed considerable experimental progress in recent years, with several physical platforms that have demonstrated the control of a large number of entangled qubits that is challenging to simulate with general purpose classical methods~\cite{Arute2019, wu2021, Kim2023, Bluvstein2024}. Despite this progress, the current error rates still greatly limit the performances and the applications of quantum computers. Quantum error correction provides a path towards the realization of large scale quantum computations with arbitrary low errors, at the price of a significant overhead in terms of number of qubits and control electronics~\cite{Campbell2017}. Recent experiments have provided fundamental demonstrations of the building blocks of quantum error correction~\cite{google2023, google2025, Putterman2025, Bluvstein2024, Krinner2022, Marques2022, Andersen2020, Sivak2023}. Thus, it can be expected that in the following years large quantum processors with lower and lower error rates will gradually be built. 

In the long path towards a fully fault-tolerant quantum computer, it is natural to ask whether many small quantum computers can be used to simulate a quantum computation on a larger one, possibly with the aid of classical resources~\cite{bravyi2016, cutqc, lostaglio2021, eddins2022}. In this manuscript, we focus on techniques that achieve this goal without the need of a quantum link between the quantum processors, which go under the name of \emph{circuit cutting} (or \emph{knitting}). Typically, the goal of these techniques is to find decompositions of quantum operations acting on a system composed of subsystems A and B, into operations that act locally only on the subsystems A and B, respectively, possibly with the aid of additional ancillas.  These techniques are usually divided into wire cutting (or time-like cuts)~\cite{peng, harada2024, pednault2023, brenner2023, lowe2023, dimatteo2022, Perlin_2021, chen2022}, in which one cleverly uses decompositions of the identity channel to split the quantum circuit, and gate cutting (or space-like cuts), where an entangling gate between subsystems A and B is directly decomposed into local operations~\cite{piveteausutter2024, schmittpiveteau2024, mitaraifujii2019, mitaraifujii2021a, mitaraifujii2021b, ufrecht2023, ufrecht2024}. Once these decompositions are obtained, the quantum operation of interest can be simulated using the standard quasiprobability sampling method~\cite{piveteausutter2024}, which is essentially the same method used in error mitigation techniques, such as probabilistic error cancellation~\cite{temme2017, endo2018, caiqem}. In fact, circuit cutting can also be interpreted as an error mitigation technique, especially when a device has limited connectivity, since cutting a gate between qubits that are far away, could lead to lower errors, compared to implementing the gate by means of SWAP operations~\cite{yamamoto2023}. Recently, several works have also shown the experimental realization of different aspects of quantum circuit cutting~\cite{yamamoto2023, nakamura2024, ufrecht2023, chen2022, suchara2020, suchara2021, pan2023, egger2024}.

The price to pay for circuit cutting is a sampling overhead that is quantified by the $1$-norm of the decomposition, i.e., the sum of the absolute values of the coefficients. This justifies the search for decompositions that minimize the $1$-norm. It has been shown that the minimum $1$-norm is related to the robustness of entanglement of the quantum operation of interest~\cite{vidal1999, harrow2025, piveteausutter2024}. If multiple quantum operations are cut independently, the total overhead scales multiplicatively, i.e., it is given by the product of the respective $1$-norms, giving rise to an exponential scaling with the number of cuts. Recently Ref.~\cite{harrow2025} has constructed a general protocol, called the double Hadamard test, which, instead, achieves a submultiplicative scaling of the sampling overhead, and applied it to further improve the application of circuit cutting to clustered Hamiltonian simulation, originally studied in Ref.~\cite{peng}. A complementary line of research has also explored optimal ways of placing wire and gate cuts in a given quantum circuit, taking into account hardware constraints~\cite{brandhofer2024, martonosi2022, chen2023, chenhansen2023}. Finally, we point out that whether circuit cutting will find application for large quantum computations is still under debate, since the sampling overhead might still be impractical for practical applications as remarked in a recent study Ref.~\cite{yangmurali2024}. 

\begin{figure}[]
    \centering
    \includegraphics[width= 1 \columnwidth]{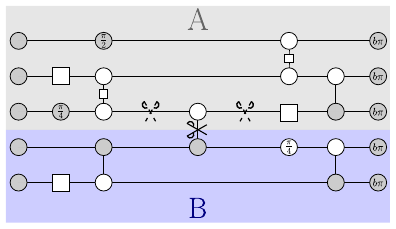}
    \caption{ ZX-diagram of an exemplary five-qubit circuit illustrating different cutting techniques. The subsystems A and B are only connected by a CNOT gate. Conventional wire cutting can be used to split the circuit by inserting two wire cuts at the location of the dashed scissors. The solid scissor represents our contribution in this paper and corresponds to applying a wire cut to the wire connecting the two qubits involved in the CNOT in the ZX-diagram representation. This effectively allows us to obtain a gate cut and this idea can potentially be applied to arbitrary ZX-diagrams representing larger unitaries.
    }
    \label{fig:example_circuit}
\end{figure}

A prerequisite for cutting a generic quantum operation is the ability to find a suited decomposition into local and implementable operations. In this manuscript, we show how decompositions of large unitary gates can be obtained using the diagrammatic language of ZX-calculus~\cite{coecke2011, Coecke_Kissinger_2017, vandewetering2020}. The application of ZX-calculus to circuit cutting has already been explored in Ref.~\cite{ufrecht2023} for the particular case of the multi-qubit controlled-Z (MCZ) gate. Our main contribution is the realization that wire cutting can be used to cut gates, as long as the wire in question is a wire in a ZX-diagram. In fact, the very notion of time- or space-like wires is lifted in ZX-diagrams since they are a special implementation of so-called \emph{string} diagrams~ ~\cite{Coecke_Kissinger_2017,vandewetering2020}.
A basic visualization of this idea is shown in Fig.~\ref{fig:example_circuit}, where the gate to be cut is a CNOT. We apply our idea to the MCZ gate on $n$-qubits, obtaining a decomposition that achieves the optimal $1$-norm of $\gamma=3$ for arbitrary $n$, which uses a single ancilla qubit without post-selection. This improves on the result reported in Ref.~\cite{ufrecht2023}, showing that $\gamma=3$ can be achieved for any $n$, at the price of a single ancilla qubit. Furthermore, we show that despite the fact that the optimal $1$-norm of an individual wire cut is $3$, one can still obtain decompositions with lower $1$-norms. In particular, we show this for a two-qubit $ZZ$-rotation with arbitrary angle $\theta$. We obtain a $1$-norm of $1 + 2 | \sin(\theta)|$, which matches the result of Ref.~\cite{mitaraifujii2021a}. The decompositions are simply obtained by simplifying the obtained ZX-diagrams using the known ZX-diagrammatic rules and carefully keeping track of the coefficients. Importantly, despite the fact that we use wire cutting protocols that require classical communication, the corresponding gate decompositions do not require it, owing to the simplifications obtained via the diagrammatic rules. A decomposition requiring classical communication means that the parts of the circuit that it decomposes are required to exchange classical information. For example, making a wire cut with classical communication means to allow the state that is re-initialized to depend on the measurement result at the cut.

The paper is organized as follows. Sec.~\ref{sec:wczx} provides preliminaries, as well as an introduction to ZX-calculus and to wire cutting. In particular, the wire-cutting protocol we will employ is described in Sec.~\ref{subsec:wire_with_comm}. The decompositions obtained by using wire cutting in ZX-diagrams are presented in Sec.~\ref{sec:zxgatecut}. In particular, Sec.~\ref{subsec:mcz} contains the improved decompositions for the MCZ gate, while Sec.~\ref{subsec:2_rz} focuses on the two-qubit $ZZ$-rotation. Furthermore, in Sec.~\ref{subsec:cutcontrolsequence} we consider a sequence of controlled unitaries that share the same control qubit and show how to separate the control qubit from the target qubits with a single wire cut. Appendix~\ref{app:smpm} discusses how the unphysical maps used in the paper can be simulated. We conclude with Sec.~\ref{sec:conc} providing a discussion and an outlook.  

\section{Wire cutting in ZX-calculus}
\label{sec:wczx}
\subsection{Preliminaries}
We start by briefly introducing the necessary notation and definitions. We consider an $n$-qubit quantum system with Hilbert space $\mathcal{H}_n = (\mathbb{C}^{2})^{\otimes n}$ and let $\mathrm{L}_{\mathrm{b}}(\mathcal{H}_n)$
denote the set of linear and bounded operators on $\mathcal{H}_n$. Also, we denote by $\mathrm{D} (\mathcal{H}_n) \subset \mathrm{L}_{\mathrm{b}}(\mathcal{H}_n) $ the set of density matrices associated with the Hilbert space $\mathcal{H}_n$. Given two operators $A, B \in \mathrm{L}_{\mathrm{b}}(\mathcal{H}_n)$ we define the Hilbert-Schmidt inner product as $\bbrakket{A}{B} = \mathrm{Tr}(A^{\dagger} B)$. The space $\mathrm{L}_{\mathrm{b}}(\mathcal{H}_n)$ equipped with the Hilbert-Schmidt inner product forms a $4^n$-dimensional Hilbert space. Thus, selecting a basis for this space, we can represent any $A\in \mathrm{L}_{\mathrm{b}}(\mathcal{H}_n)$ as a $4^n$-dimensional column vector $\kket{A}$, while we denote by $\bbra{A}$ its associated conjugate transpose. Accordingly, given this Hilbert space structure, any linear map $\mathcal{M}: \mathrm{L}_{\mathrm{b}}(\mathcal{H}_n) \mapsto \mathrm{L}_{\mathrm{b}}(\mathcal{H}_n)$ can be represented as a matrix in a specified basis, which we denote as $\hat{\mathcal{M}}$. For a unitary $U$ acting on $\mathcal{H}_n$, we denote by $\mathcal{U}$ its channel representation, that acts on an operator $A$ as $\mathcal{U}(A) = U A U^{\dagger}$, and by $\hat{\mathcal{U}}$ its representation as a Liouville superoperator. Let $X, Y$ and  $Z$ be the single-qubit Pauli operators and $I$ the single-qubit identity. The Pauli group on $n$ qubits $\mathcal{P}_n$ is the group of all possible $n$-fold tensor products of Pauli operators together with sequential multiplication as the group operation.
An element of $\mathcal{P}_n$ can be written as $i^{a} P_1 \otimes \dots \otimes P_n$ with $P_k \in \{I, X, Y, Z\}$, $\forall k \in [n]$, and $a \in \{0, 1, 2, 3 \}$. Let $\mathcal{P}_n^{(+)}$ denote the set of elements of $\mathcal{P}_n$ with $a=0$. We call the elements of $\mathcal{P}_n^{(+)}$ the $n$-qubit Pauli operators. The $n$-qubit Pauli operators form an orthogonal basis for $\mathrm{L}_{\mathrm{b}}(\mathcal{H}_n)$, but they are not normalized under the Hilbert-Schmidt inner product. Thus, we define the normalized Paulis as $\overline{P} = P/\sqrt{2^n}$ for every $P \in \mathcal{P}_n^{(+)}$. Using these definitions the $n$-qubit identity channel $\mathcal{I}$ can be decomposed in superoperator form as
\begin{equation}
    \hat{\mathcal{I}} = \sum_{P \in \mathcal{P}_n^{(+)}} \kket{\overline{P}} \bbra{\overline{P}}  = \frac{1}{2^n}\sum_{P \in \mathcal{P}_n^{(+)}} \kket{P} \bbra{P}.
    \label{eq:id_ptm_n}
\end{equation}
When expressed in the normalized Pauli basis, we call the matrix $\hat{\mathcal{M}}$ the Pauli transfer matrix (PTM). 

\begin{figure*}
    \centering
    \includegraphics[width=1 \textwidth]{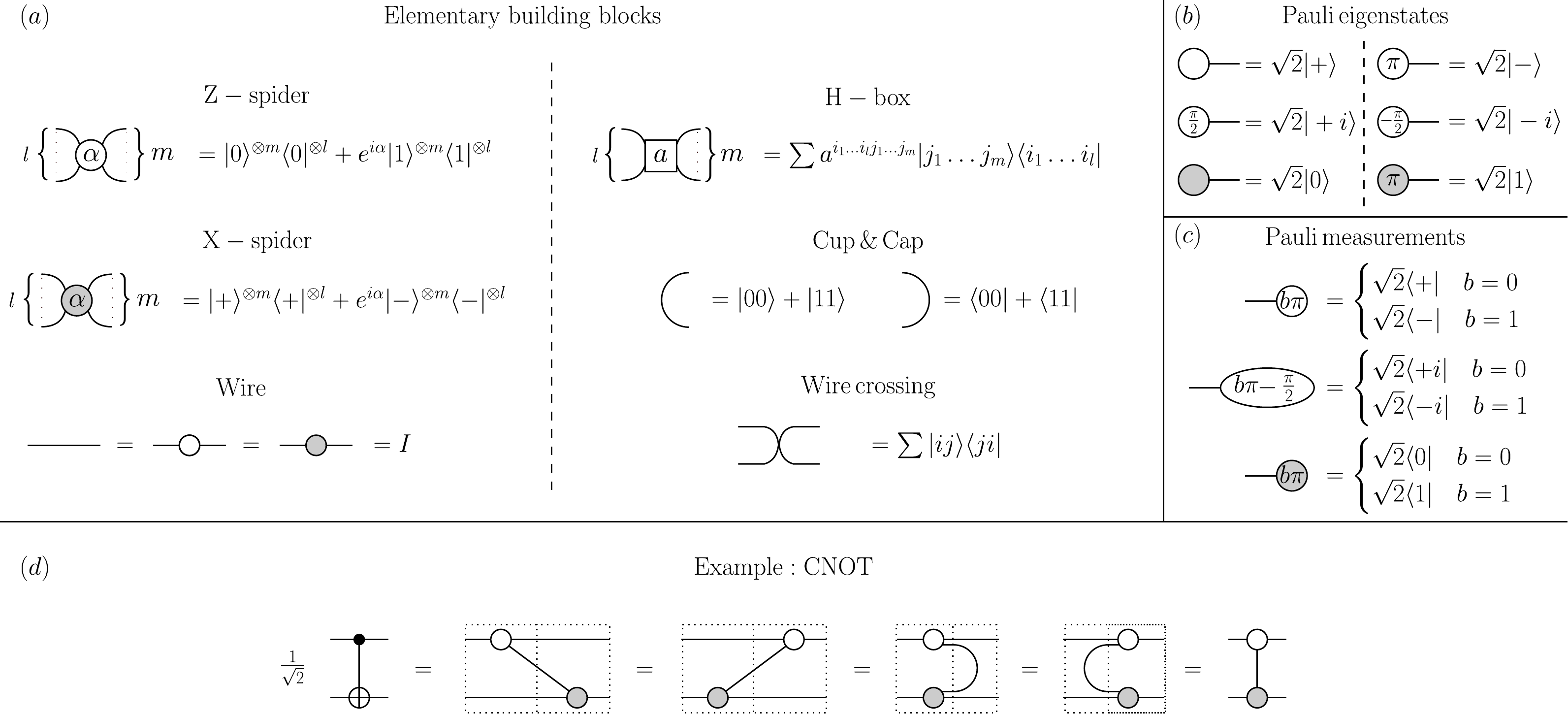}
    \caption{Fundamental ZX-diagrams used in this manuscript. (a) Z- and X-spiders are the main building blocks that characterize ZX-diagrams. The H-box generalizes the Hadamard gate, while Cup and Cap represent Bell states and Bell measurements, respectively. Wires and wire crossing represent the single-qubit identity and the SWAP gate, respectively. (b) Pauli eigenstates are proportional via $\sqrt{2}$ factors to either Z- or X-spiders with a single output and no inputs. (c) Similarly, Pauli measurements are represented via ZX-diagrams with one input and no outputs, and introducing an ancilla bit $b=0,1$ associated with the measurement outcome. (d) Representations of the CNOT gate in ZX-calculus. 
    }
    \label{fig:zx_build_block}
\end{figure*}

We are interested in expressing a quantum channel $\mathcal{E}$, i.e., a completely positive trace-preserving (CPTP) map, into a linear combination of $M$ maps $\mathcal{F}_{\nu}$ with $\nu \in \{1, \dots, M \}$. In Liouville superoperator formalism this translates to
\begin{equation}
\label{eq:generaldeco}
    \hat{\mathcal{E}} = \sum_{\nu=1}^M q_\nu \hat{\mathcal{F}}_\nu,
\end{equation}
with $q_\nu \in \mathbb{R}$. The maps $\mathcal{F}_\nu$ are the operations we allow on hardware and are not necessarily CPTP maps, as also unphysical, but simulable, maps may be allowed (see Ref.~\cite{piveteausutter2024} for a discussion). We denote a decomposition as in Eq.~\eqref{eq:generaldeco} as a set \begin{equation}
\label{eq:set_deco}
\Omega_{\mathcal{E}} := \{(q_\nu, \mathcal{F}_\nu) \, | \, \nu \in \{1, \dots, M \} \}.
\end{equation} 
By gathering the coefficients $q_\nu$ in an $M$-dimensional vector $\vec{q}$, we define the 1-norm of the decomposition $\gamma \left(\Omega_{\mathcal{E}} \right)$ as
\begin{equation}
    \label{eq:def1norm}
    \gamma\left(\Omega_{\mathcal{E}} \right) := \lVert \vec{q} \rVert_1 = \sum_{\nu=1}^M |q_\nu|.
\end{equation}
If the decompositions $\Omega_{\mathcal{E}}$ are restricted to be part of a set $S_{\mathcal{E}}$ we define the coefficient $\gamma_{\mathrm{opt}}(S_{\mathcal{E}})$ as the minimum $1$-norm over all the allowed and valid decompositions:
\begin{equation}
    \gamma_{\mathrm{opt}} (S_{\mathcal{E}}) := \mathrm{min} \, \{ \gamma(\Omega_{\mathcal{E}}) \, | \, \Omega_{\mathcal{E}} \in S_{\mathcal{E}} \}.
\end{equation}
In circuit cutting we are typically interested in restricting the maps $\mathcal{F}_\nu$ to those that can be implemented locally on two quantum systems A and B, and potentially additional ancilla qubits. Once we have a decomposition as in Eq.~\eqref{eq:generaldeco} the channel $\hat{\mathcal{E}}$ can be simulated using the standard quasiprobability sampling method \cite{piveteausutter2024}, which is also the base for quantum error mitigation techniques, such as probabilistic error cancellation \cite{temme2017}. This essentially amounts to rewriting Eq.~\eqref{eq:generaldeco} as
\begin{equation}
\label{eq:generaldeco_prob}
    \hat{\mathcal{E}} = \gamma(\Omega_{\mathcal{E}}) \sum_{\nu=1}^M p_\nu \mathrm{sign}(q_\nu) \hat{\mathcal{F}}_\nu,
\end{equation}
where the coefficients $p_\nu$ form a probability distribution and are defined as 
\begin{equation}
    \label{eq:prob}
    p_\nu = \frac{|q_\nu|}{\gamma(\Omega_{\mathcal{E}})}.
\end{equation}
If we are interested in obtaining the average of an observable $O = O_{A} \otimes O_{B}$ with initial state $\rho = \rho_{A} \otimes \rho_{B}$, i.e., 
\begin{equation}
    \langle O \rangle = \bbra{O} \hat{\mathcal{E}} \kket{\rho},
\end{equation}
an unbiased estimator $y_{O}$ for $\langle O \rangle$ can be obtained as 
\begin{equation}
    y_{O \nu} = \gamma(\Omega_{\mathcal{E}}) \mathrm{sign}(q_\nu) y_{O, \mathcal{F}_\nu},
\end{equation}%
where $y_{O, \mathcal{F}_\nu}$ is an unbiased estimator for $\bbra{O} \hat{\mathcal{F}}_\nu \kket{\rho}$. If $\hat{\mathcal{F}}_\nu$ is itself a CPTP map the estimator is merely the measurement outcome upon measurement of $O$ on the state $\hat{\mathcal{F}}_\nu \kket{\rho}$. Nonetheless, as we explain below, more general maps can be allowed, as long as, no additional sampling overhead is incurred. If this is the case, the standard argument based on Hoeffding's inequality (see Ref.~\cite{caiqem} for instance) states that the sampling overhead of the cutting protocol is $\gamma(\Omega_{\mathcal{E}})^2$. 

Following Ref.~\cite{harada2024} we define generalized measure and prepare maps $\mathcal{E}_{\mathrm{gmp}}$ on $n$ qubits as maps that act on a generic operator $A$ as
\begin{equation}
\label{eq:mpmap}
    \mathcal{E}_{\mathrm{gmp}}(A) := \sum_{\nu=1}^M a_{\nu} \mathrm{Tr}[E_{\nu} A] \rho_{\nu},
\end{equation}
where $a_{\nu}= \pm 1$, $E_{\nu} \ge 0$ are the elements of a positive operator valued measurement (POVM) satisfying $\sum_{\nu=1}^M E_{\nu} = I$, and $\rho_{\nu} \in \mathrm{D}(\mathcal{H}_n)$ are density matrices. As pointed out in Ref.~\cite{harada2024}, the definition of measure and prepare maps in Eq.~\eqref{eq:mpmap} is a generalization of the one in Ref.~\cite{horodecki2003}, which is restricted to the case $a_{\nu}=1$. In fact, only in this case the map in Eq.~\eqref{eq:mpmap} represents a physical CPTP map. We highlight that the Liouville superoperator representation of $\mathcal{E}_{\mathrm{gmp}}$ reads
\begin{equation}
\label{eq:empsup}
    \hat{\mathcal{E}}_{\mathrm{gmp}} = \sum_{\nu=1}^M a_{\nu} \kket{\rho_{\nu}}\bbra{E_{\nu}}.
\end{equation}

Finally, we also define generalized Kraus maps $\mathcal{E}_{\mathrm{gK}}$ as maps whose action on a generic operator $A$ can be written as
\begin{equation}
\label{eq:gkmap}
    \mathcal{E}_{\mathrm{gK}} (A) = \sum_{\nu=1}^M a_{\nu} K_{\nu} A K_{\nu}^{\dagger},
\end{equation}
where $a_{\nu} = \pm 1$ and the operators $K_{\nu}$ satisfy $\sum_{\nu=1}^M K_{\nu}^{\dagger} K_{\nu} = I$ \footnote{Note that in principle one can view generalized measure and prepare maps as a particular case of generalized Kraus maps, but we decided to distinguish them since measure and prepare maps do not necessarily need ancillas.}. Similarly to generalized measure and prepare maps, generalized Kraus maps are CPTP maps only if all $a_{\nu} =1$, and, in this case, Eq.~\eqref{eq:gkmap} would be one of the possible Kraus representations  of the quantum channel. As we show in Appendix~\ref{app:smpm}, even when the coefficients $a_{\nu}$ are allowed to be $-1$, both generalized measure and prepare maps and generalized Kraus maps can be effectively simulated with no additional sampling overhead. The idea is to simply keep track of the signs of the $a_{\nu}$ coefficients depending on some measurements results, which for the case of generalized Kraus maps involve ancilla qubits. 

\subsection{ZX-calculus: a mini course}
ZX-calculus is a diagrammatic language specifically designed to describe quantum processes \cite{coecke2011, Coecke_Kissinger_2017, vandewetering2020}. Each diagram in ZX-calculus is a tensor where, in general,  $l$ indices are identified as inputs and $m$ as outputs. Each index is associated with a ``leg" in the diagram. When the input legs of one diagram are connected to the output legs of another diagram the corresponding indices are contracted, i.e., the indices are summed over all possible values. Thus, a ZX-diagram is a set of connected elementary diagrams that form a specific type of tensor network. In what follows, we give a concise description of the diagrams and the combination rules that are needed to understand the derivations of the circuit cutting protocols described in Sec~\ref{sec:zxgatecut}. 

The elementary building blocks we employ in this manuscript to construct ZX-diagrams are shown in Fig.~\ref{fig:zx_build_block}a. Z- and X-spiders represent the main ingredients of ZX-calculus and are diagonal in the Z- and X-basis, respectively. Note that when the parameter $\alpha$ is not shown in a spider it is assumed to be zero. Single wires simply represent the single-qubit identity gate and can also be viewed as either a Z- or an X-spider with one input and one output and $\alpha=0$. 
The H-box 
can be viewed as a generalization of the Hadamard gate $H= \frac{1}{\sqrt{2}} \begin{pmatrix}
    1 & 1 \\
    1 & -1
\end{pmatrix}$. In fact, the H-box for $a=-1$ and $m=l=1$ is equal to $\sqrt{2} H$. For $a=-1$ we omit the label in the H-box. We note that the H-box is not independent of the other diagrams, but it is helpful to concisely represent multi-qubit controlled operations that we discuss in Sec.~\ref{sec:zxgatecut}. In fact, it is well known that the Hadamard gate can be expressed in several ways using only Z- and X-spiders (see Eq.~(45) in Ref.~\cite{vandewetering2020} for instance). H-boxes together with Z-spiders are also the main ingredients of the so-called ZH-calculus \cite{zh_calculus, vandewetering2020}.  The Cup and Cap diagrams represent Bell state preparation and Bell measurement, respectively. They are fundamental in obtaining one important property of ZX-diagrams that is at the heart of all our protocols:
\begin{center}
    \emph{Only connectivity matters in ZX-diagrams.}
\end{center}
This fact is exemplified in Fig.~\ref{fig:zx_build_block}d  where different, but equivalent, representations of the CNOT gate in ZX-calculus are shown. 
Finally, wire crossing represents a SWAP gate. Wire crossing can also be obtained only using Z- and X-spiders, but it is convenient to introduce an explicit symbol for it. 

Fig.~\ref{fig:zx_build_block}b shows how the single-qubit stabilizer states, i.e., the eigenstates of the Pauli $X$, $Y$ and $Z$ operators can be represented in ZX-calculus. Note that it is common in the ZX-calculus literature to omit proportionality factors (in this case $\sqrt{2}$), but we choose to keep them as they will be helpful in later derivations. Fig.~\ref{fig:zx_build_block}c shows how Pauli measurements can be represented in ZX-calculus. The parameter $b$ is a bit $b=0,1$ that encodes the measurement outcome.

\begin{figure*}
    \centering
    \includegraphics[width=1 \textwidth]{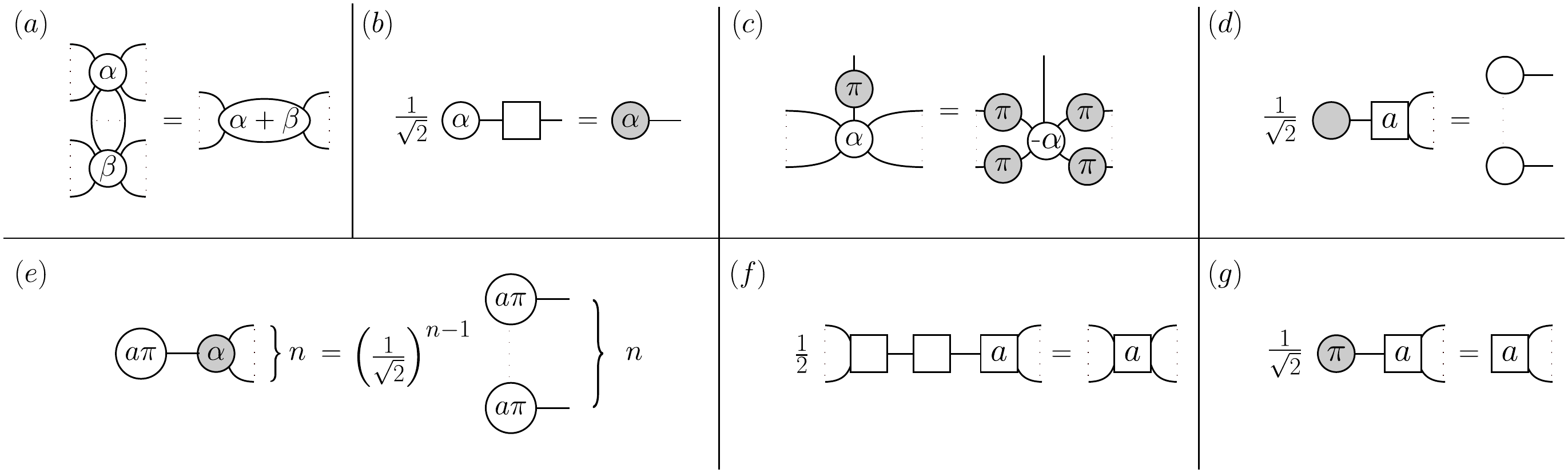}
    \caption{ZX diagrammatic rules used in this manuscript.
    }
    \label{fig:zx_rules}
\end{figure*}

An appealing feature of ZX-calculus is that diagrams can be combined using several diagrammatic rules that allow one to build intuition about quantum circuits. A summary of the diagrammatic rules used in this manuscript is shown in Fig.~\ref{fig:zx_rules}.  

\subsection{Wire cutting and its ZX-representation}
\label{sec:wc}
In this section, we give an introduction to wire cutting, which is the fundamental primer of all the cutting protocols we analyze in this manuscript. We also discuss how wire cutting can be represented using ZX-diagrams.

In what follows, we proceed by first discussing wire cutting protocols without classical communication in Sec.~\ref{subsec:wire_no_comm}. Afterwards, we explain how allowing classical communication can decrease the overhead of wire cutting in Sec.~\ref{subsec:wire_with_comm}.

\subsubsection{Without classical communication}\label{subsec:wire_no_comm}

\begin{table}[ht]
\centering
\begin{tabular}{||c c c c c||} 
 \hline
 $P$ & $a_{P0}$ & $\rho_{P0}$ &  $a_{P1}$ & $\rho_{P1}$   \\ [0.5ex] 
 \hline\hline
 $I$ & $1$ & $\ket{+i}\bra{+i}$ & $1$ & $\ket{-i}\bra{-i}$   \\ 
 \hline
 $X$ & $1$ & $\ket{+}\bra{+}$ & $-1$ & $\ket{-}\bra{-}$   \\ 
 \hline
  $Y$ & $1$ & $\ket{+i}\bra{+i}$ & $-1$ & $\ket{-i}\bra{-i}$   \\ 
 \hline
  $Z$ & $1$ & $\ket{0}\bra{0}$ & $-1$ & $\ket{1}\bra{1}$   \\ 
 \hline
\end{tabular}
\caption{Eigenstates of the single-qubit Pauli operators.}
\label{tab:eigenstates}
\end{table}

As a basic example of wire cutting, let us consider a single-qubit circuit. We assume that the qubit is initialized in a state $\kket{\rho_{\text{in}}}$. Then we apply a quantum circuit that is described by subsequent applications of the unitaries $U_{1}$ and $U_{2}$. Finally, assume that we are interested in estimating the expectation value of an observable $O$ with eigenvalues $\lambda_O \in [-1, 1]$, i.e., $O$ can be one of the Pauli operators for instance. The expectation value of $O$ is given by 
\begin{equation}
    \braket{O} = \mathrm{Tr}[O \mathcal{U}_2 \circ \mathcal{U}_1(\rho_{\mathrm{in}})] = \bbra{O} \hat{\mathcal{U}}_{2} \hat{\mathcal{U}}_{1} \kket{\rho_{\text{in}}}. \label{eq:O_uncut}
\end{equation}
We place the cut between the application of $U_{1}$ and $U_{2}$. Considering Eq.~\eqref{eq:id_ptm_n} for the single-qubit case $n=1$, and inserting it into Eq.~\eqref{eq:O_uncut}, we obtain
\begin{equation}
\label{eq:opauliid}
    \langle O \rangle = \frac{1}{2} \sum_{P \in \mathcal{P}_1^{(+)}} \bbra{O} \hat{\mathcal{U}}_2 \kket{P} \bbra{P} \hat{\mathcal{U}}_1 \kket{\rho_{\mathrm{in}}}
\end{equation}
Furthermore, we expand the vectorized single-qubit Pauli operators in their eigenbasis as
\begin{equation}
    \kket{P} =  \sum_{\mu=0}^{1} a_{P \mu} \kket{\rho_{P \mu}},   \label{eq:pauli_eigenstates}
\end{equation}
where $\rho_{P \mu}$ denotes the single-qubit stabilizer states, i.e., the eigenstates of the single-qubit Pauli operators $P$, while $a_{P \mu}$ are the corresponding eigenvalues (see Table~\ref{tab:eigenstates} for a summary). Additionally,  there is a freedom in the choice of the eigenstates for $P=I$. We choose to use the eigenstates of the Pauli $Y$ here. Using Eq.~\eqref{eq:pauli_eigenstates} we can re-write Eq.~\eqref{eq:opauliid} as
\begin{equation}
    \braket{O} = \frac{1}{2} \sum_{P \in \mathcal{P}_1^{(+)}} \sum_{\mu=0}^1 a_{P \mu} \bbra{O} \hat{\mathcal{U}}_2 \kket{\rho_{P\mu}} \bbra{P} \hat{\mathcal{U}}_1 \kket{\rho_{\mathrm{in}}}.
\end{equation}
We have effectively represented the identity as a linear combination of eight different measure and prepare channels $\mathcal{E}_{P \mu}$ of the form
\begin{equation}
\label{eq:wcemp}
   \hat{\mathcal{E}}_{P \mu}:= \kket{\rho_{P \mu}} \bbra{P} = \sum_{\nu =0}^1 a_{P \nu} \kket{\rho_{P \mu}} \bbra{\rho_{P \nu}}.
\end{equation}
Explicitly
\begin{equation}
\label{eq:impdeco}
    \hat{\mathcal{I}} = \sum_{P \in \mathcal{P}_1^{(+)}} \sum_{\mu =0}^1 q_{P \mu} \hat{\mathcal{E}}_{P \mu},
\end{equation}
with coefficients 
\begin{equation}
\label{eq:qpmu}
q_{P \mu} = \frac{a_{P \mu}}{2}.
\end{equation}
This decomposition was first introduced in the context of circuit cutting by Peng et al. in Ref.~\cite{peng}. 
As in the general case of Eq.~\eqref{eq:set_deco}, we denote the decomposition in Eq.~\eqref{eq:impdeco} as a set 
\begin{equation}
\label{eq:pengdeco}
   \Omega_{\mathcal{I}}^{(\mathrm{nCC})} =\left \{ \left(q_{P\mu}, \mathcal{E}_{P \mu} \right) \, | \, P \in \mathcal{P}_1^{(+)}, \, \mu \in \{0, 1\} \right \}.
\end{equation}
Here the superscript $(\mathrm{nCC})$ encodes the fact that this decomposition does not use classical communication. We can interpret the $q_{P \mu}$ coefficients as a quasiprobability distribution. In fact, $\sum_{P \in \mathcal{P}_1^{(+)}} \sum_{\mu=0}^1 q_{P \mu} =1$, but since some of the coefficients are negative the $1$-norm is larger than $1$: 
\begin{equation}
    \gamma \left (\Omega_{\mathcal{I}}^{(\mathrm{nCC})} \right) = \sum_{P \in \mathcal{P}_1^{(+)}} \sum_{\mu =0}^1 | q_{P \mu } | = 4.
\end{equation}
Note that $\hat{\mathcal{E}}_{P \mu}$ satisfies the definition given in Eq.~\eqref{eq:empsup} of a measure and prepare map in Liouville superoperator representation. Thus, we obtain
\begin{equation}
\label{eq:O_cut_quasiprob}
      \braket{O}  
    = \sum_{P \in \mathcal{P}_1^{(+)}} \sum_{\mu =0}^1 q_{P \mu} \bbra{O} \hat{\mathcal{U}}_2 \hat{\mathcal{E}}_{P \mu}\hat{\mathcal{U}}_1 \kket{\rho_{\mathrm{in}}}.
\end{equation}
The cutting protocol works by standard quasiprobability simulation in which each of the eight measure and prepare maps $\hat{\mathcal{E}}_{P \mu}$ is sampled with probability $p_{P \mu} = |q_{P \mu}|/\gamma \left (\Omega_{\mathcal{I}}^{(\mathrm{nCC})} \right) =1/8$, and simulated using the general protocol described in Appendix~\ref{app:smpm}. The sampling overhead is quantified by $\gamma \left (\Omega_{\mathcal{I}}^{(\mathrm{nCC})} \right)^2 = 16$. If we apply this protocol to independently cut wires on $n$ qubits in parallel, then the overhead would be $16^n$.  We remark that the measure and prepare channels $\mathcal{E}_{P \mu}$ defined in Eq.~\eqref{eq:wcemp} do not require classical communication since the prepared states $\rho_{P\mu}$ do not depend on the POVM outcome $\mu$. However, this protocol is sub-optimal even in the single-qubit case and lower sampling overheads can be achieved if we allow classical communication ~\cite{harada2024, brenner2023}. 

\subsubsection{With classical communication}
\label{subsec:wire_with_comm}
Allowing for classical communication can decrease the overhead of wire cutting. In particular, the optimal 1-norm of cutting $n$ wires with classical communication can be shown to be $2^{n+1}-1$~\cite{brenner2023}. An explicit protocol for cutting with this 1-norm is given in Ref.~\cite{brenner2023}. However, this protocol requires ancilla qubits. This requirement can be removed for the special case of parallel wire cutting~\cite{harada2024, pednault2023}, which includes the scenario of cutting a single wire. As we only perform cuts for single wires in this paper, we focus on this case in the following.

We proceed by introducing the protocol for wire cutting that we use throughout this paper. Our protocol is heavily inspired by the protocols introduced in Ref.~\cite{harada2024} and Ref.~\cite{pednault2023}. The basic idea is to group some of the $\mathcal{E}_{P \mu}$ channels together to obtain a new measure and prepare channel, while leaving the other invariant. We choose to group together all terms that are associated with the identity $I$ and the Pauli $Y$ operators. We define the measure and prepare channel
\begin{equation}
\label{eq:ep}
    \hat{\mathcal{E}}_{P} := \kket{\rho_{P 0}}\bbra{\rho_{P 0}} + \kket{\rho_{P 1}}\bbra{\rho_{P 1}},
\end{equation}
for $P \in \{X, Y, Z\}$. In particular, we include into the decomposition $\hat{\mathcal{E}}_{Y}$, while we keep all the other channels as in the Peng decomposition in Eq.~\eqref{eq:pengdeco}. The single-qubit identity can be decomposed as 
\begin{equation}
\label{eq:sdeco}
    \hat{\mathcal{I}} = q_Y \hat{\mathcal{E}}_{Y}+ \sum_{P \in \{X, Z \} } \sum_{\mu =0}^1 q_{P \mu} \hat{\mathcal{E}}_{P \mu},
\end{equation}
where $q_Y = 1$ and the $q_{P \mu}$ coefficients are the same as those for the corresponding terms in the Peng decomposition defined in Eq.~\eqref{eq:qpmu}. Thus, the decomposition is
\begin{equation}
   \Omega_{\mathcal{I}}^{(\mathrm{CC})} := \{(q_Y, \mathcal{E}_Y) \} 
   \cup
   \left \{  \left(q_{P\mu}, \mathcal{E}_{P \mu} \right) \, | \, P \in \{X, Z\}, \, \mu \in \{0, 1\} \right \},
\end{equation}
with $1$-norm
\begin{equation}
    \gamma \left(\Omega_{\mathcal{I}}^{(\mathrm{CC})} \right) = |q_Y| + \sum_{P \in \{X, Z\}} \sum_{\mu = 0}^1 |q_{P  \mu}|=  3.
\end{equation}
Here the superscript indicates that the decomposition uses classical communication. The protocol for cutting using the decomposition in Eq.~\eqref{eq:sdeco} is summarized in Table~\ref{tab:cut_wire_comm}.  Throughout the paper we call the operation where we measure in the basis $\{ \ket{+i}, \ket{-i} \}$ and re-intialize the corresponding eigenstates the ``$Y$-term". Similarly, we define the  ``$X$-term"  \ and the  ``$Z$-term" accordingly. We clearly see that in our protocol we require classical communication for the $Y$-term, as we re-initialize the eigenstates that we measured. However, we do not require classical communication for the $X$-term and the $Z$-term. We will find later that this is a useful property when attempting to cut a gate by cutting a wire in ZX-calculus. The protocols suggested in Refs.~\cite{harada2024, pednault2023} require classical communication for the $X$-, $Y$- and $Z$-term and this is the reason why we do not use them. We note that we could have grouped the terms differently and the role of the $Y$-term could have been be taken by the $X$- or the $Z$-term. When cutting the MCZ gate and the multi-qubit $Z$-rotation gate via a wire cut we will see that the choice of the term for which we apply classical communication can indeed be crucial.

In the last column of Table~\ref{tab:cut_wire_comm} we show the ZX-diagrams corresponding to the different measure and prepare maps in the protocol. 

\begin{table*}[ht]
\centering
\def\arraystretch{1.5}
\begin{tabular}{||c c c c c c c ||} 
 \hline
 Map & CPTP & $q$ & Measurement basis  & Result & Re-initialize  &  ZX-diagram \\ [0.5ex] 
 \hline\hline
   $\mathcal{E}_{Y}$ & Yes & +1
   & $\{ \ket{+i},\ket{-i} \} $ & $\ket{+i}$ 
    &   $\ket{+i}$   & 
 $
 \frac{1}{2}
\begin{ZX}
 \zxNone{} \rar & \zxFracZ-{\pi}{2} & \zxFracX-{\pi}{2} \rar & \zxNone{}
\end{ZX}
$
 \\ 
     & &  &
   & $\ket{-i}$   &     $\ket{-i}$ 
        &  $
 \frac{1}{2}
\begin{ZX}
 \zxNone{} \rar & \zxFracZ{\pi}{2} & \zxFracX{\pi}{2} \rar & \zxNone{}
\end{ZX}
$ \\
 \hline
  $\mathcal{E}_{X0}$ & No &$\frac{1}{2}$ & $\{ \ket{+},\ket{-} \}$& $\ket{+}$
&$\ket{+}$ 
&    $
 \frac{1}{2}
\begin{ZX}
 \zxNone{} \rar & \zxZ{} & \zxZ{} \rar & \zxNone{}
\end{ZX}
$
\\ 
& &  &  &$\ket{-}$ 
&  $\ket{+}$ 
&  $
 \frac{1}{2}
\begin{ZX}
 \zxNone{} \rar & \zxZ{\pi} & \zxZ{} \rar & \zxNone{}
\end{ZX}
$
   \\ 
\hline

  $\mathcal{E}_{X1}$ & No & $-\frac{1}{2}$ & $\{ \ket{+},\ket{-} \}$& $\ket{+}$
&$\ket{-}$ 
&    $\frac{1}{2}
\begin{ZX}
 \zxNone{} \rar & \zxZ{} & \zxZ{\pi} \rar & \zxNone{}
\end{ZX}
$
\\ 
& &  &  &$\ket{-}$ 
&  $\ket{-}$ 
&  $
 \frac{1}{2}
\begin{ZX}
 \zxNone{} \rar & \zxZ{\pi} & \zxZ{\pi} \rar & \zxNone{}
\end{ZX}
$ \\
\hline 

  $\mathcal{E}_{Z0}$ & No & $\frac{1}{2}$ & $\{ \ket{0},\ket{1} \}$& $\ket{0}$
&$\ket{0}$ 
&    $
 \frac{1}{2}
\begin{ZX}
 \zxNone{} \rar & \zxX{} & \zxX{} \rar & \zxNone{}
\end{ZX}
$
\\ 
& &  &  &$\ket{1}$ 
&  $\ket{0}$ 
&  $
 \frac{1}{2}
\begin{ZX}
 \zxNone{} \rar & \zxX{\pi} & \zxX{} \rar & \zxNone{}
\end{ZX}
$
   \\ 
\hline

  $\mathcal{E}_{Z1}$ & No & $-\frac{1}{2}$ & $\{ \ket{0},\ket{1} \}$& $\ket{0}$
&$\ket{1}$ 
&    $
  \frac{1}{2}
\begin{ZX}
 \zxNone{} \rar & \zxX{} & \zxX{\pi} \rar & \zxNone{}
\end{ZX}
$
\\ 
& & &  &$\ket{1}$ 
&  $\ket{1}$ 
&  $
 \frac{1}{2}
\begin{ZX}
 \zxNone{} \rar & \zxX{\pi} & \zxX{\pi} \rar & \zxNone{}
\end{ZX}
$ \\

 \hline

\end{tabular}
\caption{ Protocol for cutting a wire with classical communication for the $Y$-term. The last column gives ZX-diagrams corresponding to the different channels. }
\label{tab:cut_wire_comm}
\end{table*}

\section{Gate cutting in ZX-calculus}
\label{sec:zxgatecut}

\subsection{Cutting the MCZ gate}
\label{subsec:mcz}

\begin{figure}[]
    \centering
    \includegraphics[width= 1 \columnwidth]{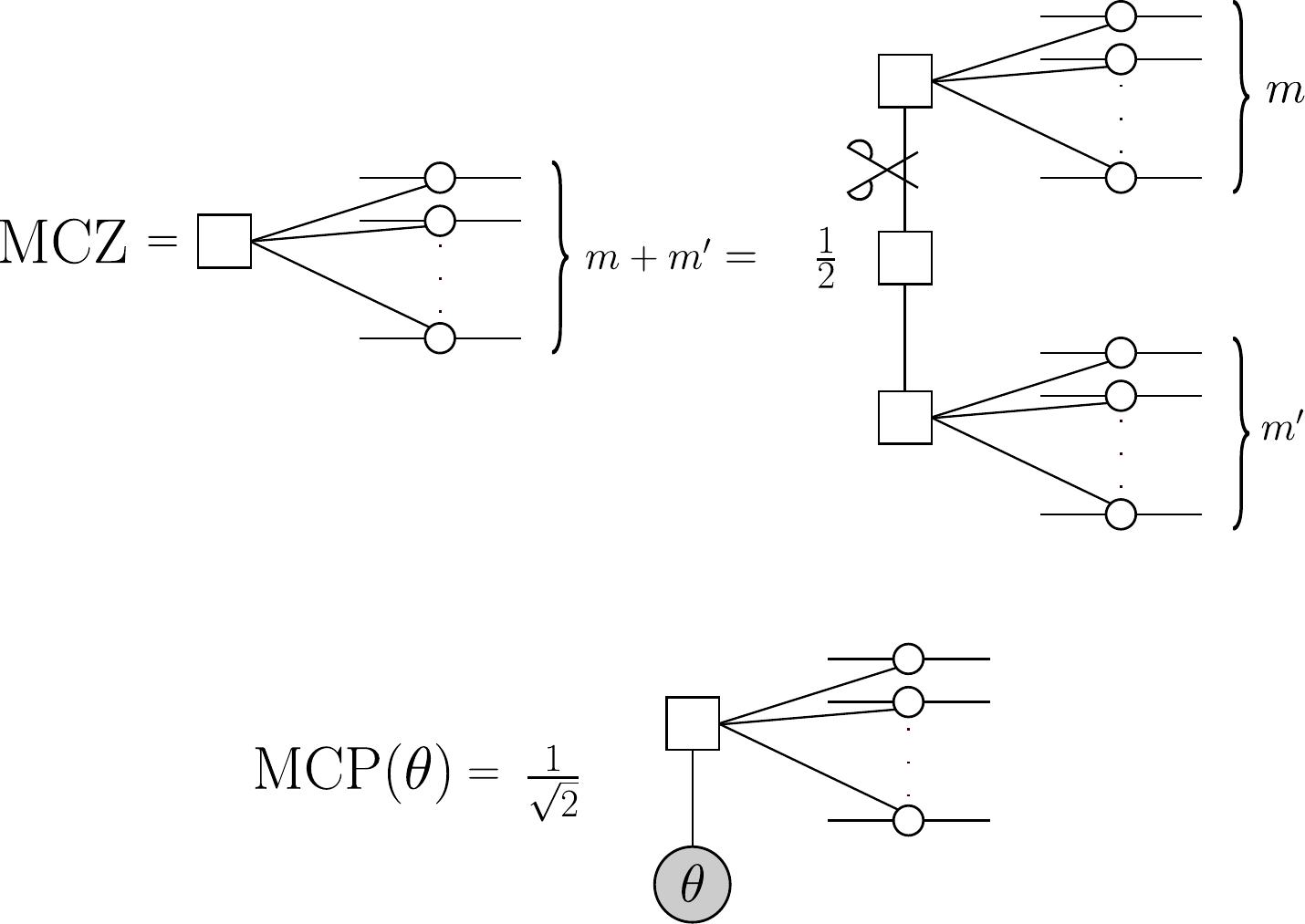}
    \caption{(Top) Representation of an MCZ gate acting on $n=m+m'$ qubits in ZX-calculus. On the right-hand side the ZX-diagram is rewritten using the H-box fusion rule in Fig.~\ref{fig:zx_rules}f. The scissor indicates where a single wire cut can be inserted to cut the MCZ gate. (Bottom) Representation of a multi-controlled phase gate $\mathrm{MCP}(\theta)$ in ZX-calculus.
    }
    \label{fig:mcz_gate}
\end{figure}

\begin{figure*}[]
    \centering
    \includegraphics[width=1 \textwidth]{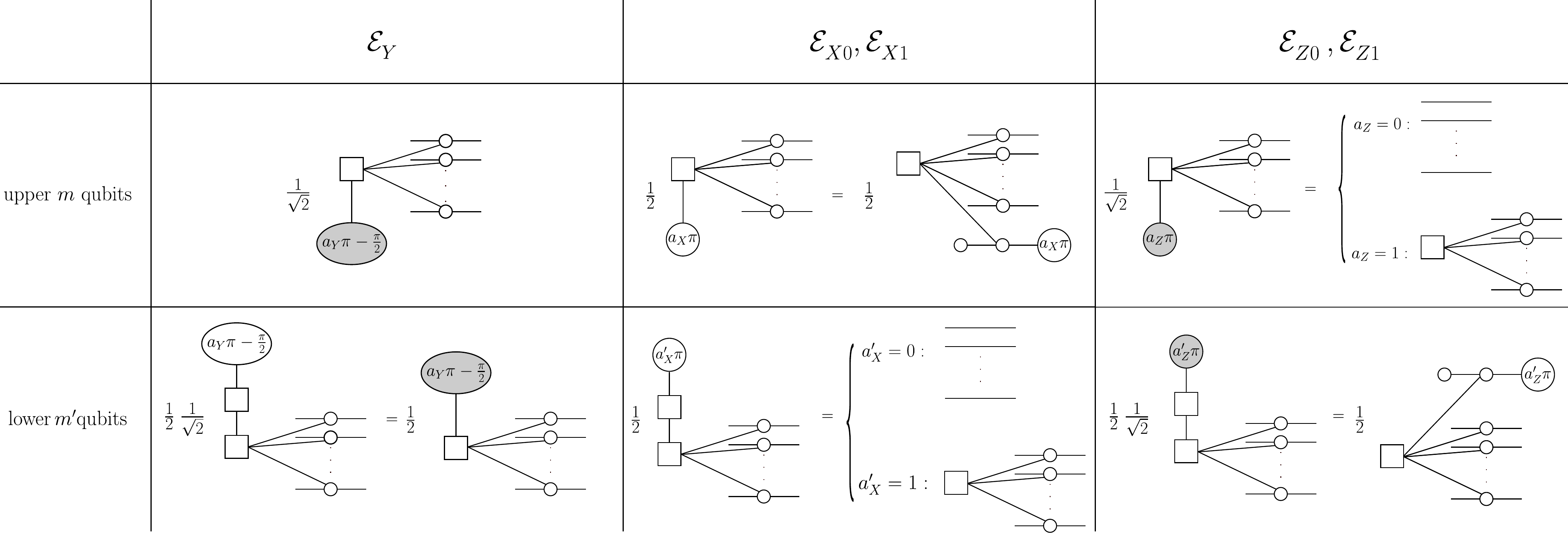}
    \caption{ZX-diagrams resulting from cutting a wire in an MCZ gate. The variables $a_{Y},a_{X}$, $a_{X}'$, $a_{Z}$, $a_{Z}'$ can take values in $\{ 0,1 \}$. The diagrammatic rules shown in Fig.~\ref{fig:zx_rules} are used to convert the ZX-diagrams to valid quantum circuits. Note that every diagram is correctly normalized apart from the one for the $Y$-term acting on the lower $m'$ qubits which requires an additional factor of $\sqrt{2}$. This means that we need to multiply the probability to sample from the correctly normalized diagram by $1/2$, as explained in the main text.
    }
    \label{fig:mcz_deco}
\end{figure*}

In the following, we show how to cut an MCZ gate by cutting a single wire in ZX-calculus. The MCZ gate often appears in quantum algorithms and it is particularly relevant in the framework of quantum signal processing~\cite{gilyen2019, martyn2021, dalzell2023}. The MCZ gate is defined as $\mathrm{MCZ} = \mathrm{diag}(1, \dots, 1, -1)$ and can be represented in ZX-calculus with the help of an H-box as shown in Fig.~\ref{fig:mcz_gate}(top). Additionally, we define the multi-controlled phase gate $\mathrm{MCP}(\theta)=\mathrm{diag}(1, \dots, 1, e^{i \theta})$, whose representation as a ZX-diagram is shown in Fig.~\ref{fig:mcz_gate}(bottom). Note that $\mathrm{MCZ} = \mathrm{MCP}(\pi)$. We remark that the synthesis of the $\mathrm{MCP}(\theta)$ gate in terms of elementary two-qubit gates has been studied using ZX-calculus in Ref.~\cite{staudacher2024}. 

Assume that the gate acts on $n=m+m'$ qubits. We cut the gate into circuits that act locally on the upper $m$ or the lower $m'$ qubits. For this purpose, similarly to Ref.~\cite{ufrecht2023}, we exploit the H-box fusion rule (see Fig.~\ref{fig:zx_rules}) to re-write the MCZ gate with three H-boxes, as shown on the right-hand  side of Fig.~\ref{fig:mcz_gate}. This representation of the MCZ gate as ZX-diagram is particularly well suited for our approach of cutting as we just need to cut a single wire to cut the gate. The position at which we insert the wire cut is denoted with a scissor. We split this section into two parts. In Sec.~\ref{subsubsec:mcz_der} we derive the protocol for cutting the MCZ gate. Afterwards, we discuss the protocol in Sec.~\ref{subsubsec:mcz_dis}. 

\subsubsection{Derivation of the cutting protocol}\label{subsubsec:mcz_der}

Let us now cut the MCZ gate by cutting the wire at the scissor in Fig.~\ref{fig:mcz_gate}. We do this by inserting the different ZX-diagrams from Table~\ref{tab:cut_wire_comm}. We obtain different kinds of diagrams that act locally on the lower or the upper qubits and show them in Fig.~\ref{fig:mcz_deco}. Besides, we show there how to adapt these diagrams such that they correspond to quantum circuits. The diagrams contain the variables $a_{Y}$, $a_{X}$, $a_X'$, $a_{Z}$, $a_{Z}'$ which can take the values $\{0,1\}$. The spiders that carry these variables correspond to the two eigenstates of the Pauli $Y,X,$ and $Z$.

In the derivation of the protocol, it is important to correctly take into account the scalars that are associated with each ZX-diagram. In fact, the diagrams that we get by cutting a wire in an MCZ gate carry a scalar of $1/4 = (1/2) \cdot (1/2)$. One factor of $1/2$ comes from the H-box fusion rule. The other factor of $1/2$ originates from the wire cut, as can be seen in Table~\ref{tab:cut_wire_comm}. For the $X$- and $Z$-term, we split the overall scalar $1/4$ between the diagrams for the lower and the upper qubits so that the diagrams are properly normalized. Instead, for the $Y$-term only the diagram for the upper qubits is properly normalized while the diagram for the lower qubits shall be multiplied by a scalar of $\sqrt{2}$ for proper normalization. The effect of this scalar can be taken into account in the protocol by multiplying the coefficients associated with the diagrams that originate from the $Y$-term by $1/2$. In the following paragraphs we discuss how to interpret and implement the diagrams shown in Fig.~\ref{fig:mcz_deco} to obtain a decomposition of the MCZ gate. 

The $Y$-term gives two unitary contributions, which in quantum channel representation are
\begin{equation*}
    \mathcal{MCP}^{(m)}\left(\frac{\pi}{{2}}  \right) \otimes \mathcal{MCP}^{(m')}\left(\frac{\pi}{{2}}  \right)
\end{equation*}
and 
\begin{equation*}
    \mathcal{MCP}^{(m)}\left(-\frac{\pi}{{2}}  \right) \otimes \mathcal{MCP}^{(m')}\left(-\frac{\pi}{{2}}  \right).
\end{equation*}
Interestingly, these terms do not require neither classical communication nor measurements, despite the fact that the $Y$-term in the original cutting protocol does involve both. Because of the scalar factor discussed above the coefficient associated with both of these operations is given by $q_Y/2 = 1/2$. 

The $X$-term gives rise to two contributions. In the lower $m'$ qubits either the identity $\mathcal{I}^{\otimes m'}$ or a MCZ $\mathcal{MCZ}^{(m')}$ is applied. The map implemented on the upper $m$ qubits instead is not a CPTP map, but rather a generalized Kraus map as defined in Eq.~\eqref{eq:gkmap}. This map consists in the application of an MCZ gate between the $m$ qubits in the upper register  and an ancilla qubit prepared in the $\ket{+}$ state. Subsequently, the ancilla is measured in the $X$-basis and depending on whether the result is $\ket{+}$ or $\ket{-}$, the estimator of the observable of interest is multiplied by $+1$ or $-1$, respectively (see Appendix~\ref{app:smpm}). We denote this generalized Kraus map as $\mathcal{E}_{\mathrm{MCZ}-\mathrm{M}_{X} }^{(m)}$ and its action on an $m$-qubit density matrix $\rho$ is defined as
\begin{multline}
\label{eq:mapmczmeas}
   \mathcal{E}_{\mathrm{MCZ}-\mathrm{M}_{X} }^{(m)} (\rho) = \mathrm{Tr}_a\left(\Pi_{a+}^{(m)}\mathcal{MCZ}^{(m + 1)}(\rho \otimes \ket{+}\bra{+}) \right) - \\
\mathrm{Tr}_a\left(\Pi_{a-}^{(m)}\mathcal{MCZ}^{(m + 1)}(\rho \otimes \ket{+}\bra{+})\right),
\end{multline}
where $\Pi_{a +}^{(m)} = I^{\otimes m} \otimes \ket{+}\bra{+}$, $\Pi_{a -}^{(m)} = I^{\otimes m} \otimes \ket{-}\bra{-}$, and $\mathrm{Tr}_a(\cdot)$ denotes the partial trace over the ancilla qubit. As we see from Fig.~\ref{fig:mcz_deco} the $Z$-term gives rise to exactly the same kinds of terms, but with $m$ and $m'$ inverted. The corresponding $q$ coefficients coincide with those of the $X$- and $Z$-terms in Table~\ref{tab:cut_wire_comm}. The full decomposition $\Omega_{\mathrm{MCZ}}$ is summarized in Table~\ref{tab:mcz_deco}, and $\forall m, m'$  such that $n=m + m'$ its $1$-norm is the same as the one of the wire-cutting protocol we used~\footnote{We note that we also obtained a decomposition of the $\mathrm{MCP}(\theta)$ gate for any $\theta$ with 1-norm equal to $3$, but this decomposition is sub-optimal for general $\theta$. While not included in the manuscript, the decomposition can be found in the github repository associated with this paper (see Data Availability section)}:
\begin{equation}
\label{eq:gammamcz}
    \gamma\left(\Omega_{\mathrm{MCZ}} \right) = \gamma\left(\Omega_{\mathcal{I}}^{(\mathrm{CC})} \right) =  3.
\end{equation}
This decomposition is optimal for any partition of the $n$ qubits into $m$ and $m'$ sub-registers. In fact, Ref.~\cite{schmittpiveteau2024} (see remark 7.1) shows that the minimum $1$-norm for the three-qubit MCZ gate, i.e. the controlled-controlled-Z (CCZ) gate is $3$. The optimality for the $n$-qubit MCZ follows by noticing that by setting $n-3$ qubits in $\ket{1}$ we can use it to obtain a CCZ, and thus the optimal 1-norm for the CCZ is a lower bound for the optimal 1-norm of the $n$-qubit MCZ.

\begin{table}[ht]
\centering
\begin{tabular}{||c c c||} 
 \hline
 Map & CPTP & $q$   \\ [0.5ex] 
 \hline\hline
   $\mathcal{MCP}^{(m)}\left(\frac{\pi}{{2}}  \right) \otimes \mathcal{MCP}^{(m')}\left(\frac{\pi}{{2}}  \right)$ & Yes & $\frac{1}{2}$ 
 \\ 
   $\mathcal{MCP}^{(m)}\left(-\frac{\pi}{{2}}  \right) \otimes \mathcal{MCP}^{(m')}\left(-\frac{\pi}{{2}}  \right)$ & Yes & $\frac{1}{2}$ 
 \\ 
 $  \mathcal{E}_{\mathrm{MCZ}-\mathrm{M}_{X} }^{(m)}  \otimes \mathcal{I}^{\otimes m'} $ & No & $\frac{1}{2}$ 
 \\
  $ \mathcal{E}_{\mathrm{MCZ}-\mathrm{M}_{X} }^{(m)} \otimes \mathcal{MCZ}^{( m')} $ & No & $-\frac{1}{2}$ 
 \\
 $\mathcal{I}^{\otimes m} \otimes   \mathcal{E}_{\mathrm{MCZ}-\mathrm{M}_{X} }^{(m')} $ & No & $\frac{1}{2}$ 
 \\
  $\mathcal{MCZ}^{(m)}\otimes \mathcal{E}_{\mathrm{MCZ}-\mathrm{M}_{X} }^{(m')} $ & No & $-\frac{1}{2}$ 
 \\

 \hline

\end{tabular}
\caption{Decomposition $\Omega_{\mathrm{MCZ}}$ of an MCZ gate acting on $n=m+m'$ qubits obtained via wire cutting and ZX-calculus.}
\label{tab:mcz_deco}
\end{table}

\subsubsection{Discussion }
\label{subsubsec:mcz_dis}

In Sec.~\ref{subsubsec:mcz_der} we derived a protocol for cutting an MCZ gate by cutting a wire in ZX-calculus. Interestingly, although we use a wire cut that requires classical communication, the resulting decomposition of the MCZ gate does not need classical communication. The reason for this is twofold. First, note that we use a wire cut that only exploits classical communication in the $Y$-term. Second, the circuits in the decomposition of the MCZ gate that come from the $Y$-term in the wire cut do not contain measurements, but only gates. We can thus ensure that we choose $a_{Y}=0$ or $a_{Y}=1$ for the circuits acting on the lower and the upper qubits without using classical communication during the execution of the cutting protocol. The circuits coming from the $X$- and the $Z$-term however do contain measurements. Thus, the choice of a suitable wire cut is crucial when cutting a gate by cutting a wire in this gate. In particular, if we chose a wire cut that requires classical communication for the $X$- or the $Z$-term the resulting cutting protocol for the MCZ gate would require classical communication. 

Our derivation of the cut of the MCZ gate is inspired by Ref.~\cite{ufrecht2023}, where the authors also cut an MCZ gate with ZX-calculus. However, there the idea to cut a gate by cutting wires was not used. In the general case of an arbitrary number of qubits in the upper and lower circuit the protocol derived in Ref.~\cite{ufrecht2023} requires a 1-norm of $\gamma=6$, which is higher than the 1-norm we obtain in Eq.~\eqref{eq:gammamcz}. However, we remark that our protocol requires the use of an ancilla qubit, while the protocol from  Ref.~\cite{ufrecht2023} does not need ancilla qubits. 

It is known that the two-qubit version of the MCZ gate, i.e., the controlled-Z (CZ) gate, can be cut with 1-norm $\gamma=3$ without ancilla qubits~\cite{mitaraifujii2021a}. This was also shown in Ref.~\cite{ufrecht2023}. This is also possible in our case. In fact, the ancilla qubits can be removed because in the case of a CZ gate all the H-boxes just come with one input and one output wire. This makes them proportional to Hadamard gates and allows us to apply more rules in ZX-calculus. We do not show the protocol explicitly since it is equivalent, up to single-qubit $Z$-rotations, to the protocol for cutting a two-qubit $ZZ$-rotation derived with ZX-calculus in Sec.~\ref{subsubsec:2rz_gamma_smaller} and also discussed in Ref.~\cite{mitaraifujii2021a, egger2024}. 

\begin{figure}[]
    \centering
    \includegraphics[width= 0.5 \columnwidth]{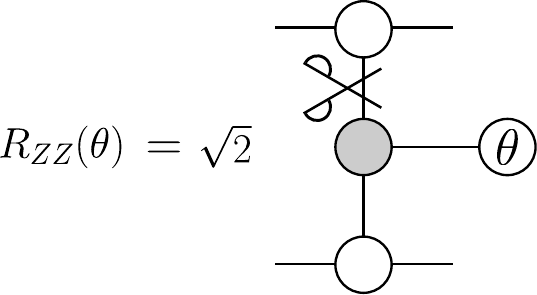}
    \caption{Representation of a two-qubit $ZZ$-rotation corresponding to the unitary  $R_{ZZ}(\theta)$ defined in Eq.~\eqref{eq:rzztheta} as a ZX-diagram. The gate is cut by cutting the wire that connects the two qubits as indicated by the scissor.
    }
    \label{fig:phase_gadget_theta}
\end{figure}

\subsection{Cutting the two-qubit $ZZ$-rotation gate}
\label{subsec:2_rz}
Here we discuss how to cut a two-qubit $ZZ$-rotation gate described by the unitary 
\begin{equation}
\label{eq:rzztheta}
R_{ZZ}(\theta) = e^{- i \frac{\theta}{2} Z \otimes Z }.
\end{equation}We note that the ability to cut this gate allows us to cut any two-qubit Pauli exponential of the form $R_{P_{1}P_{2}}(\theta) = \exp(- i \theta P_{1} \otimes P_{2}/2)$ with $P_{1}$ and $P_{2}$ arbitrary single-qubit Paulis. The reason for this is that the rotation $R_{P_{1}P_{2}}(\theta)$ can always be decomposed into a two-qubit $ZZ$-rotation and some single-qubit gates. 

The two-qubit $ZZ$-rotation gate corresponds to the ZX-diagram shown in Fig.~\ref{fig:phase_gadget_theta}. We see that we can cut the gate by cutting the wire that connects the two qubits, as indicated by the scissor. Since cutting a wire with classical communication comes with a 1-norm of $\gamma=3$, this is also the resulting 1-norm of the cut of the two-qubit $ZZ$-rotation gate. We derive the protocol for this gate cut in Sec.~\ref{subsubsec:2rz_gamma3}. We will see that it requires the use of an ancilla qubit. At the same time, it has been shown that the two-qubit $ZZ$-rotation gate can be cut with 1-norm $\gamma=1+2 |\sin( \theta)| $ and without ancilla qubits~\cite{mitaraifujii2021a, piveteausutter2024}. We find that we can reach this 1-norm if we reshuffle the channels in our protocol. In this step we also remove the need for  the ancilla qubits. We show this in Sec.~\ref{subsubsec:2rz_gamma_smaller}. For completeness, we remark that cutting the two-qubit $ZZ$-rotation gate is sufficient for cutting an $m+m'$-qubit gate of the form $\exp(- i \theta Z^{\otimes m+ m'}/2 )$~\cite{ufrecht2024}. We discuss this in Sec.~\ref{sec:mn_rz}.

\subsubsection{Cut with 1-norm $\gamma=3$}\label{subsubsec:2rz_gamma3}

\begin{figure*}[]
    \centering
    \includegraphics[width=1 \textwidth]{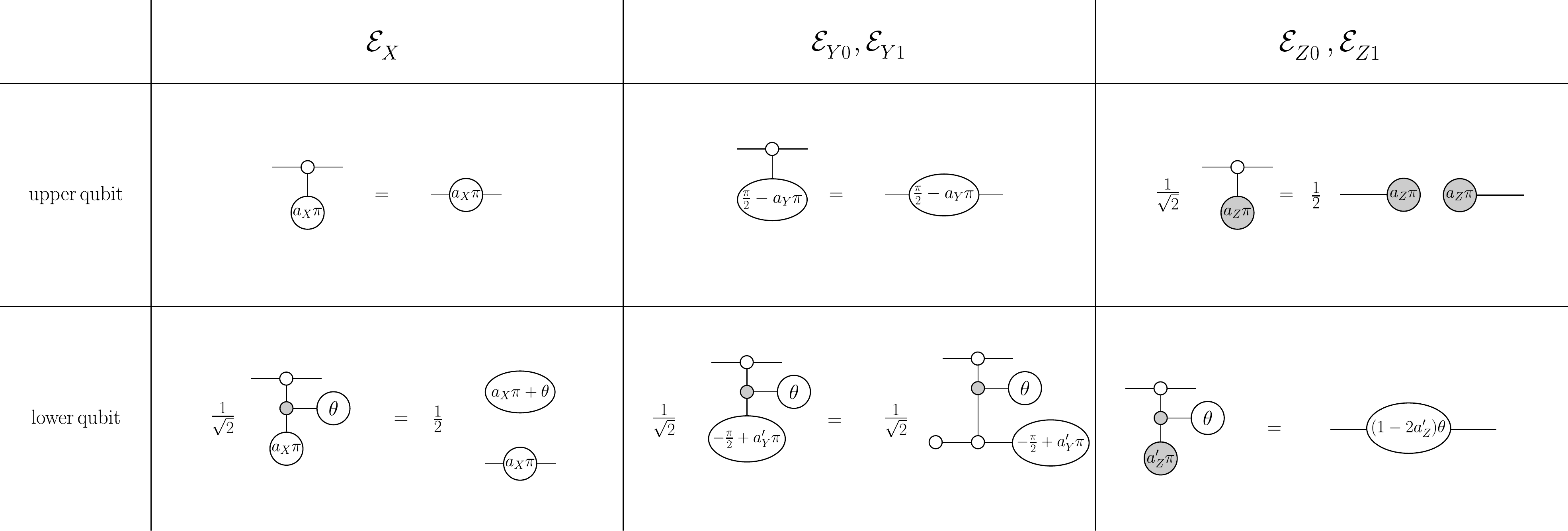}
    \caption{ZX-diagrams resulting from cutting a wire in an $R_{ZZ}(\theta)$ gate. The variables $a_{X},a_{Y}$, $a_{Y}'$, $a_{Z}$, $a_{Z}'$ can take values in $\{ 0,1 \}$. The diagram for the lower qubit and the $X$-term is not correctly normalized, and thus we need to correct the coefficient in the decomposition accordingly.
    }
    \label{fig:2rz_gate_cut}
\end{figure*}

Let us derive the cutting protocol for the two-qubit $ZZ$-rotation gate by cutting a wire in this gate at the scissor in Fig.~\ref{fig:phase_gadget_theta}. To proceed, we do use the wire cutting protocol described in Table~\ref{tab:cut_wire_comm}, but with the role of $X$ and $Y$ exchanged. Thus, we use classical communication for the $X$-term and not for the $Y$-term. In this way, the ZX-diagrams simplify in such a way that the resulting cut of the two-qubit $ZZ$-rotation gate does not require classical communication. Besides, in the wire cut we express all spiders in the $Y$-term as $Z$-spiders. The ZX-diagrams obtained by inserting this cut into the ZX-diagram associated with $R_{ZZ}(\theta)$ shown in Fig.~\ref{fig:phase_gadget_theta}, are depicted in Fig.~\ref{fig:2rz_gate_cut}. Note that to get the diagram for the $Z$-term and the lower qubit we used the fact that $R_{X}(\pi) R_{Z}(\theta) R_{X}(\pi) =R_{Z}(-\theta) $ and hence for $a_{Z}'=0$ we should apply $R_{Z}(\theta)$ and for $a_{Z}'=1$ we should apply $R_{Z}(-\theta)$. 

Note that the combined scalar of the diagrams for the upper and the lower qubit is $\sqrt{2}/2= 1/\sqrt{2}$. The reason for this is that we get a factor of $1/2$ for inserting the wire cut and the diagram of the two-qubit rotation gate comes with a factor of $\sqrt{2}$. We see that all diagrams apart from the one for the lower qubit and the $X$-term are correctly normalized. This diagram contains a $Z$-spider with entry $a_{X} \pi + \theta$ that is disconnected from the rest of the circuit. This spider represents the scalar $1+e^{i(a_{X} \pi + \theta) }$. So in total, the diagram for the lower qubit and the $X$-term requires an additional factor of $\frac{2}{1+e^{i(a_{X} \pi + \theta) }}$. We can add this factor to the diagram by multiplying the associated coefficient by 
\begin{equation*}
     \frac{1}{4} | 1+e^{i(a_{X} \pi + \theta) }|^2 = \frac{1}{2} \Bigl( 1+ \cos( a_{X} \pi + \theta) \Bigr).
\end{equation*}
We now have everything we need to give the full protocol for cutting the two-qubit $ZZ$-rotation gate. The decomposition that we denote as $\Omega_{R_{ZZ}(\theta)}^{(a)}$ is summarized in Table~\ref{tab:rzz_deco}. The map $\mathcal{E}_{R_{ZZ}(\theta)-\mathrm{M}_{Y}}$ is defined via its action on a density matrix $\rho$ in analogy with Eq.~\eqref{eq:mapmczmeas}:
\begin{multline}
   \mathcal{E}_{R_{ZZ}(\theta)-\mathrm{M}_{Y}} (\rho) = \mathrm{Tr}_a\left(\Pi_{a+i}\mathcal{R}_{ZZ}(\theta) (\rho \otimes \ket{+}\bra{+}) \right) - \\
\mathrm{Tr}_a\left(\Pi_{a-i}\mathcal{R}_{ZZ}(\theta)(\rho \otimes \ket{+}\bra{+})\right),
\end{multline}
where $\Pi_{a +i} = I \otimes \ket{+i}\bra{+i}$, $\Pi_{a -i} = I \otimes \ket{-i}\bra{-i}$. Furthermore, the measure and prepare map $\mathcal{\overline{E}}_Z$ is defined in analogy with Eq.~\eqref{eq:ep}, but with a different sign:
\begin{equation}
\label{eq:ezbar}
    \hat{\overline{\mathcal{E}}}_{Z} := \kket{\rho_{Z 0}}\bbra{\rho_{Z 0}} - \kket{\rho_{Z 1}}\bbra{\rho_{Z 1}}.
\end{equation}
The $1$-norm of this decomposition is independent of $\theta$ and given by
\begin{equation}
    \gamma \left(\Omega_{R_{ZZ}(\theta)}^{(a)} \right) = 3.
\end{equation}

\begin{table}[ht]
\centering
\begin{tabular}{||c c c||} 
 \hline
 Map & CPTP & $q$     \\ [0.5ex] 
 \hline\hline
   $\mathcal{I}  \otimes \mathcal{I}$ & Yes & $\frac{1}{2}(1+\cos (\theta))$ 
 \\ 
   $\mathcal{R}_Z\left(\pi  \right) \otimes \mathcal{R}_Z\left(\pi  \right)$ & Yes & $\frac{1}{2}(1-\cos (\theta))$ 
 \\ 
 $\mathcal{R}_{Z} \left( \frac{\pi}{2} \right) \otimes \mathcal{E}_{R_{ZZ}(\theta)-\mathrm{M}_{Y}}$ & No & $\frac{1}{2}$ 
 \\
  $\mathcal{R}_{Z} \left(-\frac{\pi}{2} \right) \otimes
  \mathcal{E}_{R_{ZZ}(\theta)-\mathrm{M}_{Y}}
   $ & No & $-\frac{1}{2}$ 
 \\
 $\mathcal{\overline{E}}_Z \otimes \mathcal{R}_Z(\theta) $ & No & $\frac{1}{2}$ 
 \\
  $\mathcal{\overline{E}}_Z \otimes \mathcal{R}_Z(-\theta)$ & No & $-\frac{1}{2}$ 
 \\

 \hline

\end{tabular}
\caption{Decomposition $\Omega_{R_{ZZ}(\theta)}^{(a)}$ of a two-qubit $\mathcal{R}_{ZZ}(\theta)$ gate into operations that only act on the upper or the lower qubit obtained via wire cutting and ZX-calculus. The $1$-norm of this decomposition is $\gamma \left(\Omega_{R_{ZZ}(\theta)}^{(a)} \right) = 3$.}
\label{tab:rzz_deco}
\end{table}

\subsubsection{Cut with 1-norm $\gamma=1+2 |\sin( \theta)| $ } \label{subsubsec:2rz_gamma_smaller}

In this section we show how to reshuffle the terms in the cutting protocol derived in Sec.~\ref{subsubsec:2rz_gamma3} to decrease the 1-norm to $\gamma=1+2 |\sin( \theta)| $, as previously obtained in Refs.~~\cite{mitaraifujii2021a, piveteausutter2024}. To intuitively explain how this works, let us first consider the trivial case $\theta=0$, that should lead to $\gamma=1$. In this case, clearly the map $\mathcal{R}_Z\left(\pi  \right) \otimes \mathcal{R}_Z\left(\pi  \right)$ has a zero coefficient. Additionally, the maps $\mathcal{R}_{Z} \left(\pm \pi/2 \right) \otimes \mathcal{E}_{R_{ZZ}(\theta)-\mathrm{M}_{Y}} $ both give a zero contribution, while the contribution of the two maps $\mathcal{\overline{E}}_Z \otimes \mathcal{R}_{Z}(\theta =0) = \mathcal{\overline{E}}_Z \otimes \mathcal{I}$ in Table~\ref{tab:rzz_deco} cancel due to the opposite sign of the coefficients. Thus, the only remaining channel in the decomposition is $\mathcal{I} \otimes \mathcal{I}$ with coefficient $1$ which equals the 1-norm, as expected. 

Let us now extend the argument of a pairwise cancellation of the maps in Table~\ref{tab:rzz_deco} to arbitrary values of $\theta$. In this case, the maps do not fully cancel but we can still decrease the 1-norm. First, we note that
\begin{equation}
   \hat{\mathcal{E}}_{R_{ZZ}(\theta)-\mathrm{M}_{Y}}   =  \sin(\theta) \hat{\overline{\mathcal{E}}}_{Z}. 
\end{equation}
Thus, we can simply incorporate the $\sin(\theta)$ in the corresponding coefficient $q$ and use $\hat{\overline{\mathcal{E}}}_{Z}$ in place of $ \hat{\mathcal{E}}_{R_{ZZ}(\theta)-\mathrm{M}_{Y}} $, removing the need of ancilla qubits. Second, we use that
\begin{multline}
    \hat{\mathcal{R}}_{Z}(\theta) - \hat{\mathcal{R}}_{Z}(-\theta) = \\
    \sin(\theta) \left(\hat{\mathcal{R}}_{Z}\left(\frac{\pi}{2} \right) - \hat{\mathcal{R}}_{Z}\left(-\frac{\pi}{2}\right)\right).
\end{multline}
Using these relations in the decomposition in Table~\ref{tab:rzz_deco}, we obtain a new decomposition that we denote as $\Omega_{R_{ZZ}(\theta)}^{(b)}$, that we summarize in Table~\ref{tab:rzz_deco_gamma_theta}. The $1$-norm of this decomposition depends on $\theta$ and is given by 
\begin{equation}
    \gamma \left(\Omega_{R_{ZZ}(\theta)}^{(b)} \right) = 1 + 2 | \sin(\theta) |.
\end{equation}
We remark that this is exactly the same protocol as given in Ref.~\cite{mitaraifujii2021a}.

\begin{table}[ht]
\centering
\begin{tabular}{||c c c ||} 
 \hline
 Map & CPTP & Coefficient $q$\\ [0.5ex] 
 \hline\hline
   $\mathcal{I}  \otimes \mathcal{I}$ & Yes & $\frac{1}{2}(1+\cos (\theta))$ 
 \\ 
   $\mathcal{R}_Z\left(\pi  \right) \otimes \mathcal{R}_Z\left(\pi  \right)$ & Yes & $\frac{1}{2}(1-\cos (\theta))$ 
 \\ 
 $\mathcal{R}_{Z} \left( \frac{\pi}{2} \right) \otimes \mathcal{\overline{E}}_Z $ & No & $\frac{1}{2} \sin(\theta)$ 
 \\
  $\mathcal{R}_{Z} \left(-\frac{\pi}{2} \right) \otimes \mathcal{\overline{E}}_Z $ & No & $-\frac{1}{2}\sin(\theta)$ 
 \\
 $\mathcal{\overline{E}}_Z \otimes \mathcal{R}_Z\left(\frac{\pi}{2} \right) $ & No & $\frac{1}{2} \sin(\theta)$ 
 \\
  $\mathcal{\overline{E}}_Z \otimes \mathcal{R}_Z\left(-\frac{\pi}{2}\right)$ & No & $-\frac{1}{2} \sin(\theta)$ 
 \\

 \hline

\end{tabular}
\caption{Decomposition $\Omega_{R_{ZZ}(\theta)}^{(b)}$ of a two-qubit $\mathcal{R}_{ZZ}(\theta)$ gate into operations that only act on the upper or the lower qubit obtained by combining the maps of the decomposition $\Omega_{R_{ZZ}(\theta)}^{(a)}$ in Table~\ref{tab:rzz_deco}. The $1$-norm of this decomposition is $\gamma \left(\Omega_{R_{ZZ}(\theta)}^{(b)} \right) = 1 + 2 | \sin(\theta) |$.}
\label{tab:rzz_deco_gamma_theta}
\end{table}

\subsubsection{Cutting the multi-qubit $Z^{\otimes n}$-rotation gate}\label{sec:mn_rz}

\begin{figure}[]
    \centering
    \includegraphics[width= 1\columnwidth]{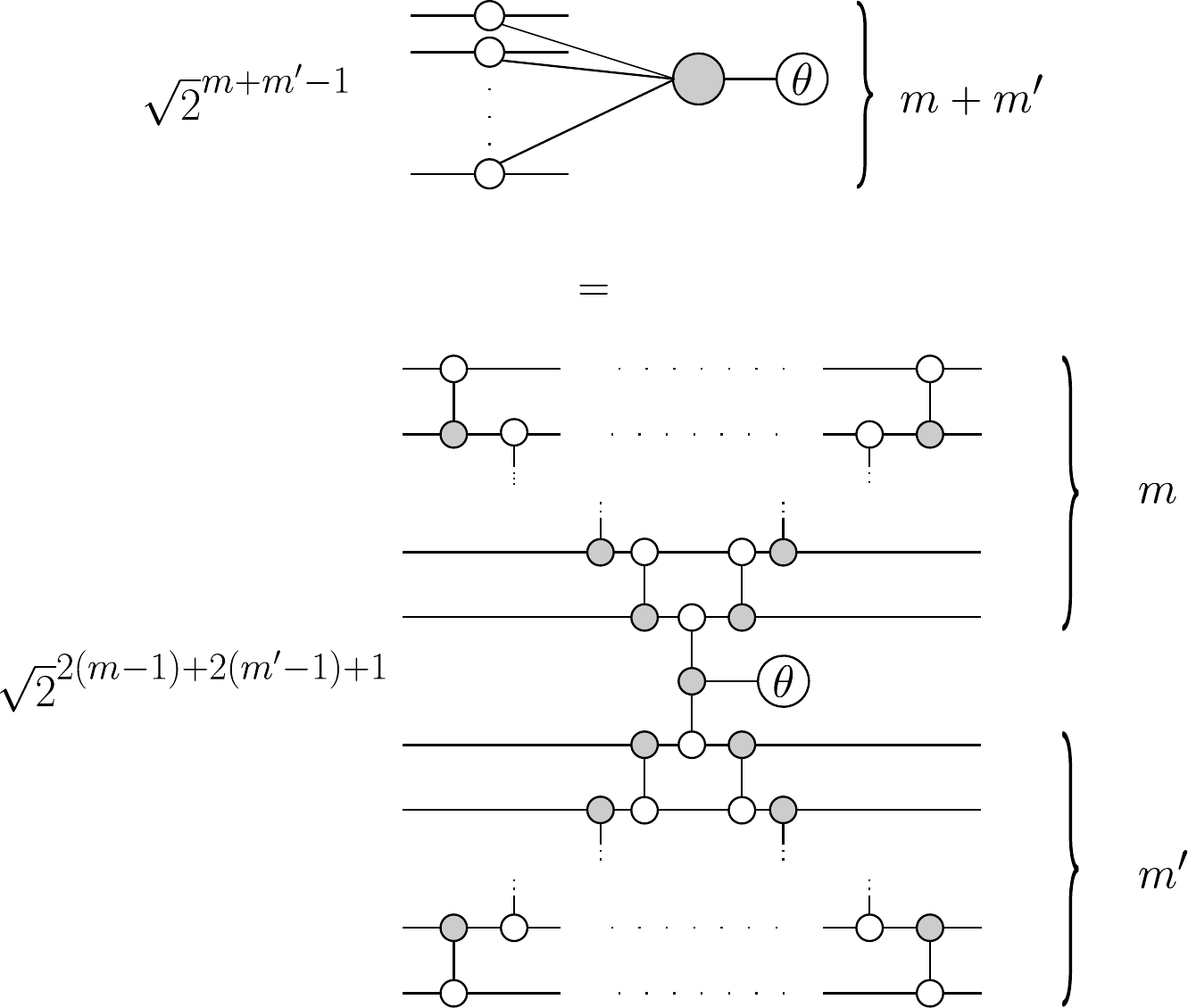}
    \caption{Representation of a multi-qubit $Z^{\otimes m + m'}$-rotation in terms of CNOT gates and a two-qubit $ZZ$-rotation~\cite{ufrecht2024, ufrecht2023}.
    }
    \label{fig:mn_rz_cnots}
\end{figure}

We remark that the multi-qubit $Z^{\otimes n}$-rotation gate, i.e., $\exp(-i \theta Z^{\otimes n}/2)$ can be cut optimally by cutting the two-qubit $ZZ$-rotation gate, as pointed out in Ref.~\cite{ufrecht2024}. The authors of Ref.~\cite{ufrecht2024} exploit the fact that a multi-qubit $Z^{\otimes n}$-rotation gate acting on $n=m+m'$ qubits can be expressed as a combination of CNOT gates and a two-qubit $ZZ$-rotation gate~\cite{mcz_cz_cut}, as shown in Fig.~\ref{fig:mn_rz_cnots}. The scalar in this diagram comes from the fact that the CNOT gates and the two-qubit $ZZ$-rotation gate require a scalar of $\sqrt{2}$ and in total we have $2 (m-1) + 2(m'-1)+1$ gates in the circuit. We note that in turn this protocol can be used to cut any Pauli exponential of the form $\exp(-i \theta P/2)$ with $P \in \mathcal{P}_n^{(+)}$, since this is equivalent to a multi-qubit $Z^{\otimes n}$-rotation up to single-qubit gates. 

\subsection{Cutting a sequence of single-qubit controlled unitaries}
\label{subsec:cutcontrolsequence}

\begin{figure}[]
\vspace{0.5cm}
    \centering
    \includegraphics[width= 1\columnwidth]{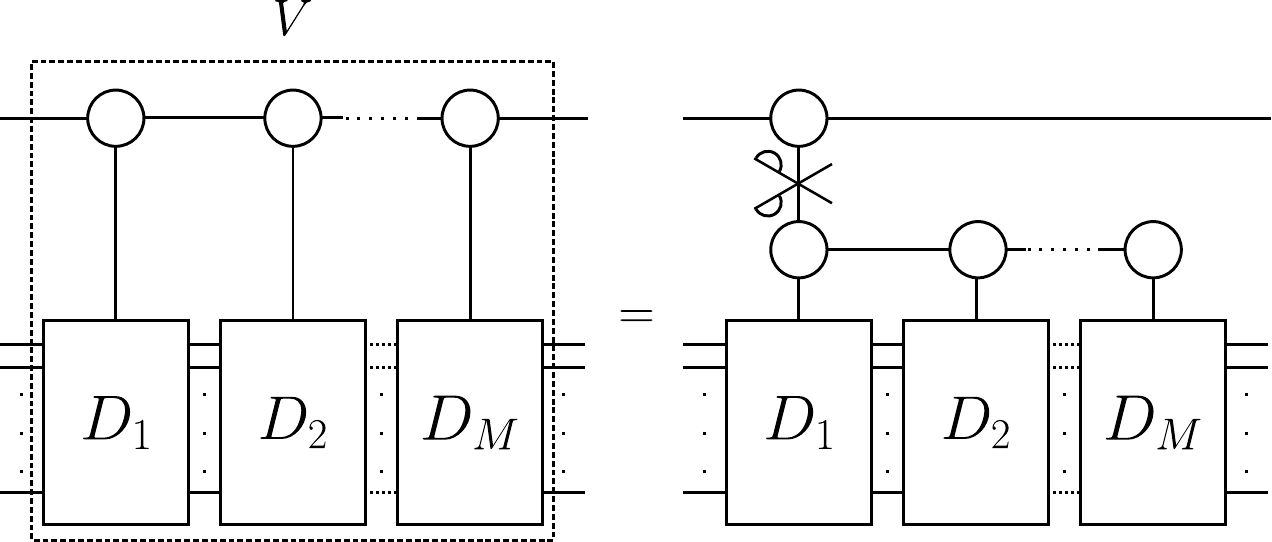}
    \caption{Sequence of $M$ single-qubit controlled unitaries with the same qubit. The boxes with label $D_1, D_2, \dots, D_M$, represent different ZX-diagram associated with each unitary. On the right the fusion rule is used to rewrite the diagram in a form that is suited for joint cutting. The scissor indicates the position of the cut.
    }
    \label{fig:cutmulticontrolled}
\end{figure}

\begin{figure*}[]
    \centering
    \includegraphics[width=1 \textwidth]{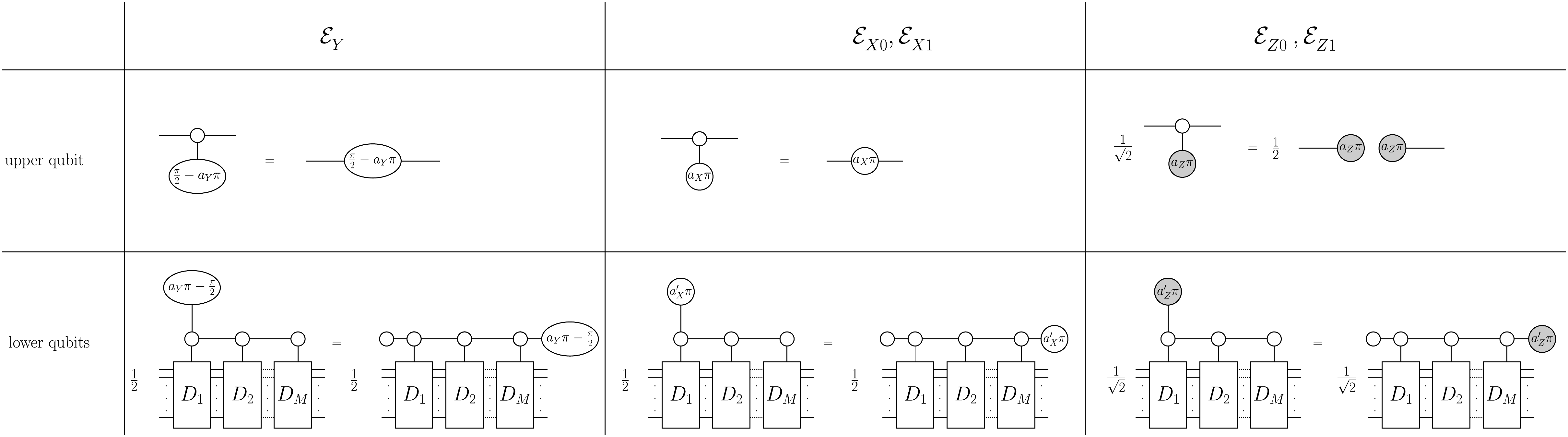}
    \caption{ZX-diagrams resulting from cutting a wire in a sequence of $M$ single-qubit controlled unitaries gate with the same qubit as control. The variables $a_{X},a_{Y}$, $a_{Y}'$, $a_{Z}$, $a_{Z}'$ can take values in $\{ 0,1 \}$. 
    }
    \label{fig:cut_deco_series_controlled}
\end{figure*}

In this section, we discuss another example of decomposition obtained using wire cutting in ZX-calculus, namely the case of a sequence of $M$ controlled unitaries with the same control qubit shown in Fig.~\ref{fig:cutmulticontrolled}. We see that we can separate the control qubit from the target qubits by cutting a single wire. We remark that to make this cut only with gate cuts or cuts of horizontal wires, we would need to apply at least two cuts. In particular, we would either have to cut the wire with the control qubit twice or have to apply a gate cut to all $M$ controlled unitaries.

In what follows, we denote the total unitary to be cut as $V$, and accordingly its associated quantum channel as $\mathcal{V}$, while $\Pi_{a s}$ denotes the projector onto the ancilla being in the state $s$. We remark that this case may be of interest in chips with limited connectivity, when the control qubit is ``far away" from the target qubits. On the other hand, as we will see, the general decomposition we obtain uses $V$ itself with an ancilla as control, so it is less interesting from a pure circuit cutting point of view. Nonetheless, further simplifications are possible depending of the specific controlled unitaries to be implemented.

Inserting the cut as denoted by the scissor on the right of Fig.~\ref{fig:cutmulticontrolled}, we obtain the diagrams shown in Fig.~\ref{fig:cut_deco_series_controlled}. To formulate the decomposition we define the map $\mathcal{E}_{R_Z V}$ coming from the $Y$-term which acts on a density matrix $\rho$ of the total system as
\begin{multline}
\mathcal{E}_{R_Z V}(\rho) = \mathrm{Tr}_a \left[\Pi_{a +i} \mathcal{R}_Z \left(\frac{\pi}{2} \right) \otimes \mathcal{V} (\rho \otimes \ket{+} \bra{+})\right] + \\
\mathrm{Tr}_a \left[\Pi_{a -i} \mathcal{R}_Z \left(-\frac{\pi}{2} \right) \otimes \mathcal{V} (\rho \otimes \ket{+} \bra{+}) \right]
\end{multline}
where the channel $\mathcal{V}$ is implemented with the ancilla as control qubit as shown in Fig.~\ref{fig:cut_deco_series_controlled}. The map $\mathcal{E}_{R_Z V}$ is a CPTP map and can be implemented by measuring the ancilla on the $Y$ basis and performing either a $\pi/2$ or a $-\pi/2$ $Z$-rotation on the top ancilla qubit, depending on the outcome of the measurement. Thus, in the general case, it requires classical communication. Additionally, we define the maps $
\mathcal{E}_{V-\mathrm{M}_{X}}$ and $ \mathcal{E}_{V-\mathrm{M}_{Z}}$ that act on a density matrix $\rho$ of the bottom register as
\begin{multline}
\mathcal{E}_{V-\mathrm{M}_{X}}
(\rho) = \mathrm{Tr}_a \left[\Pi_{a +} \mathcal{V} (\rho \otimes \ket{+} \bra{+})\right] - \\
\mathrm{Tr}_a \left[\Pi_{a -} \mathcal{V} (\rho \otimes \ket{+} \bra{+})\right],
\end{multline}
and 
\begin{multline}
\mathcal{E}_{V-\mathrm{M}_{Z}}
(\rho) = \mathrm{Tr}_a \left[\Pi_{a 0} \mathcal{V} (\rho \otimes \ket{+} \bra{+})\right] - \\
\mathrm{Tr}_a \left[\Pi_{a 1} \mathcal{V} (\rho \otimes \ket{+} \bra{+})\right],
\end{multline}
respectively. The full decomposition $\Omega_V$ is given in Table~\ref{tab:cut_deco_series_controlled} and has $1$-norm $\gamma(\Omega_V)=3$.  

\begin{table}
\centering
\begin{tabular}{||c c c||} 
 \hline
 Map & CPTP & $q$   \\ [0.5ex] 
 \hline\hline
  $\mathcal{E}_{R_Z V}$ & Yes & $1$ 
 \\
 $\mathcal{I} \otimes \mathcal{E}_{V-\mathrm{M}_{X}}$ & No & $\frac{1}{2}$ 
\\
 $\mathcal{R}_{Z}(\pi) \otimes \mathcal{E}_{V-\mathrm{M}_{X}}$ & No & $- \frac{1}{2}$ 
\\ 
 $\overline{\mathcal{E}}_{Z}\otimes \mathcal{I} \otimes 
 \mathcal{E}_{V-\mathrm{M}_{Z}}$ & No & $1$ 
\\
 \hline

\end{tabular}
\caption{Decomposition $\Omega_V$ of the sequence of controlled unitaries shown in Fig.~\ref{fig:cutmulticontrolled} obtained by combining the diagrams in Fig.~\ref{fig:cut_deco_series_controlled}. The $1$-norm of this decomposition is $\gamma(\Omega_V)=3$.}
\label{tab:cut_deco_series_controlled}
\end{table}

In the repository that accompanies this manuscript (see Data Availability section), we provide Python code that shows the correctness of the decomposition for the case of a CNOT with qubit $1$ as control and qubit $2$ as target followed by a $\mathrm{CPHASE}(\theta)$ acting on qubit $1$ and $3$.

\section{Conclusions and outlook}\label{sec:conc}
We have shown how ZX-calculus allows us to bridge wire and gate cutting of quantum circuits. In particular, gate cutting decompositions can be obtained by inserting wire cuts in the ZX-diagram associated with the gate. We applied this idea to an $n$-qubit MCZ gate, obtaining an improved decomposition with $1$-norm equal to $3$ for any $n$, and to the two-qubit $ZZ$-rotation gate, reproducing the result of Ref.~\cite{mitaraifujii2021a} using only the ZX-diagrammatic rules. Our work highlights the connection between wire and gate cutting and that ZX-calculus is a natural framework for exploring this connection. For example, it provides insights into the question of why classical communication can decrease the 1-norm of a wire cut but cannot decrease the 1-norm for cutting certain gates. In fact, we obtained gate cutting protocols without classical communication, even though we use a wire cut with classical communication.

Our work also leaves some questions open for future research. First of all, while we argue that it is always possible to insert wire cuts in ZX-diagrams, this does not necessarily mean that the corresponding diagrams represent physical, or at least simulable, operations. This is an usual and well known issue when dealing with ZX-diagrams. We have found that this is the case for the gates we analyzed, but it might not be true in general. Second, in this manuscript we have only explored the possibility of a single-wire cut. However, it is known that multiple wires can be cut more efficiently via joint cuts, instead of individual cuts~\cite{harada2024, pednault2023, brenner2023, lowe2023, piveteausutter2024}. Thus, it would be interesting to see whether gate decompositions can be obtained via inserting multi-qubit wire cuts into ZX-diagrams.

\begin{acknowledgments}
M. S. acknowledges funding from the German Federal Ministry of Education and Research (BMBF) under the program ``Quantum technologies – from basic research to
market" (Project QSolid, contract number 13N16149). M. S. was also partially funded by the Deutsche Forschungsgemeinschaft (DFG,
German Research Foundation) under Germany’s Excellence Strategy – Cluster of
Excellence Matter and Light for Quantum Computing (ML4Q) EXC 2004/1 –
390534769.

We thank Christian Ufrecht for useful comments on the manuscript.
\end{acknowledgments} 

\section*{Data availability}
The Python code to check the correctness of the decompositions is available at \url{https://github.com/cianibegood/cczx}.

\appendix 

\numberwithin{equation}{section}

\section{Simulating unphysical maps}
\label{app:smpm}
In this appendix, we explain how to simulate the generalized measure and prepare map $\mathcal{E}_{\mathrm{gmp}}$ in Eq.~\eqref{eq:mpmap} and the generalized Kraus map $\mathcal{E}_{\mathrm{gK}}$ in Eq.~\eqref{eq:gkmap} when they do not represent physical, CPTP maps, i.e., when some of the coefficients $a_{\nu}$ are equal to $-1$. The derivation will also make clear that there is no additional sampling overhead associated with these maps, despite the negative coefficients. 

Let us start by considering the generalized measure and prepare channel $\mathcal{E}_{\mathrm{gmp}}$. Without loss of generality we assume that our goal is to estimate $\langle O \rangle = \mathrm{Tr}(O \mathcal{E}_{\mathrm{gmp}}(\rho)) = \bbra{O} \hat{\mathcal{E}}_{\mathrm{gmp}}\kket{\rho}$, with $\rho \in \mathrm{D}(\mathcal{H}_n)$ a generic $n$-qubit density matrix and $O$ a Hermitian operator with eigenvalues $\lambda_O$ satisfying $\lambda_O \in [-1, 1]$~\footnote{There is no loss of generality since we can consider $O$ to be an $n$-qubit Pauli matrix  $O \in \mathcal{P}_n^{(+)}$ in which case $\lambda_O = \pm 1$, and Pauli matrices form a basis for $n$-qubit operators.}. We can construct an unbiased estimator for $\langle O \rangle$ in the following way. We simply perform the POVM with elements $E_{\nu}$ on the initial state $\rho$ and subsequently prepare the state $\rho_{\nu}$, accordingly. This essentially amounts to implementing the CPTP map associated with $\mathcal{E}_{\mathrm{gmp}}$, that is the channel obtained by setting all coefficients equal to $+1$. By measuring the observable $O$ in $\rho_{\nu}$, we obtain a sample $\tilde{y}_{O \nu}$, but in order to keep track of the sign the output of the protocol is $y_{O\nu} = a_{\nu} \tilde{y}_{O \nu}$. $y_{O \nu}$ is a random variable that provides an unbiased estimator for $\langle O \rangle$. Additionally, the range of $y_{O \nu}$ is the same as the one of $\tilde{y}_{O \nu}$, i.e., $y_{O \nu} \in [-1, 1]$. Thus, according to Hoeffding's inequality there is no additional overhead in the number of samples needed to estimate $\langle O \rangle$ with a certain accuracy and a certain probability~\cite{caiqem}. 

The argument for the simulability of the generalized Kraus map $\mathcal{E}_{\mathrm{gK}}$ follows similarly. The only additional ingredient we need is the fact that a CPTP map with Kraus operators $K_{\nu}$ can always be realized by making the system interact with an ancilla system and tracing out the ancilla~\cite{NielsenChuang}. Thus, each Kraus operator $K_{\nu}$ can be associated with a measurement performed on the ancilla system and a specific measurement outcome. By physically performing this measurement, we can keep track of the sign $a_{\nu}$ associated with $K_{\nu}$ and multiply the estimator accordingly as in the simulation of the generalized measure and prepare channel. The map $ \mathcal{E}_{\mathrm{MCZ}-\mathrm{M}_{X} }^{(m)} $ in Eq.~\eqref{eq:mapmczmeas}, provides a simple intuitive example of this process. Also in this case the range of the estimator does not change and there is no additional sampling overhead.

\bibliography{literature.bib}

\end{document}